\documentclass[useAMS,usenatbib]{mn2e}
\usepackage{graphicx}
\usepackage{txfonts}
\usepackage{natbib}
\usepackage{wrapfig}
\usepackage{longtable}
\usepackage[usenames]{color}

\include{packages}
\include{MainMacros}    

\def\arcsec {\hbox{$^{\prime\prime}$}}

\def\Real{{\rm I\mathchoice{\kern-0.70mm}{\kern-0.70mm}{\kern-0.65mm}%
  {\kern-0.50mm}R}}

\bibpunct{(}{)}{;}{a}{}{,}

\title
[Cluster galaxies die quietly]
{Star formation and AGN activity in SDSS cluster galaxies}

\author[Anja von der Linden et al.]
{\parbox[t]{\textwidth}{
Anja von der Linden$^{1,2}$
\thanks{E-mail:anja@slac.stanford.edu}, 
Vivienne Wild$^{3,2}$,
Guinevere Kauffmann$^{2}$,
Simon D. M. White$^{2}$,
Simone Weinmann$^{2}$}\\
        \vspace*{3pt}
\\
$^{1}$Kavli Institute for Particle Astrophysics and Cosmology, 
              Stanford University, 
              452 Lomita Mall,
              Stanford, CA  94305-4085, USA\\
$^{2}$Max Planck Institut f\"ur Astrophysik, 
              Karl-Schwarzschild-Str. 1, Postfach 1317,
              85741 Garching, Germany\\
$^{3}$Institut d'Astrophysique de Paris,
              UMR 7095, 
              98 bis Bvd Arago, 
              75014 Paris, France}
\begin{document}

\date{2009}

\pagerange{\pageref{firstpage}--\pageref{lastpage}} \pubyear{2009}

\maketitle

\label{firstpage}

\begin{abstract}
  We investigate the recent and current star formation activity of
  galaxies as function of distance from the cluster center in a sample
  of 521 SDSS clusters at $z<0.1$. We characterize the recent star
  formation history by the strength of the 4000\AA$\;$ break and the
  strength of the Balmer absorption lines, and thus probe the star
  formation history over the last $\sim 2$Gyr.  We show that when the
  Brightest Cluster Galaxies are excluded from the galaxy sample,
  there is no evidence for mass segregation in the clusters, so that
  differences in cluster and field populations cannot simply be
  attributed to different mass functions.  We find a marked star
  formation--radius relation in that almost all galaxies in the
  cluster core are quiescent, i.e. have terminated star formation a
  few Gyr ago. This star formation--radius relation is most pronounced
  for low-mass galaxies and is very weak or absent beyond the virial
  radius. The typical star formation rate of non-quiescent galaxies
  declines by approximately a factor of two towards the cluster
  center. However, the fraction of galaxies with young stellar
  populations indicating a recently completed starburst or a
  truncation of star formation does not vary significantly with
  radius.  These results favor a scenario in which star formation is
  quenched slowly, on timescales similar to the cluster crossing time,
  i.e. a few Gyr.  The fraction of star-forming galaxies which host a
  powerful optical AGN is also independent of clustercentric radius,
  indicating that the link between star formation and AGN in these
  galaxies operates independent of environment. The fraction of red
  galaxies which host a weak optical AGN decreases, however, towards
  the cluster center, with a similar timescale as the decline of star
  forming galaxies.
  Our results can be fully explained by a gradual decline of star
  formation rate upon infall into the cluster, 
  and rule out significant contributions from more
  violent processes at least beyond cluster radii $\gtrsim0.1
  R_{200}$.

\end{abstract}



\section{Introduction}

Galaxies in dense environments, i.e. in galaxy clusters and groups,
are observed to have different properties than galaxies in less rich
environments - their star formation activity is subdued compared to
field galaxies, and the occurrence of early-type galaxies is higher
than in the field \citep[][and references thereto]{hub36,dre80}. In
the field, star formation and morphology are not independent of each
other: low star formation rate is correlated with early-type
morphology (i.e. elliptical and S0 galaxies), whereas high star
formation activity is typical of late-type morphology (disk
galaxies). This bimodality is modified in dense environments, as
evidenced by the atypical galaxies: the occurrence of passive spiral
galaxies is higher in clusters than in the field, whereas star-forming
ellipticals are rarer in clusters \citep{bnb08}. This suggests that
star formation rate couples most strongly to environment, with
morphology being only a secondary correlation. Evidence for this has
been found in the local universe both in terms of local density
\citep{kwh04,beh05,bnb08}, as also from studies of clusters
\citep{chz05,qbb06,bay08,wkb09}.

In a hierarchical universe, clusters grow by accreting galaxies from
less dense environments. In order to form / maintain the \textit{star
  formation -- density relation}, galaxy properties thus need to be
altered upon infall into a group or cluster: star formation needs to
be suppressed, and some disk systems may need to be restructured to
create spheroidal systems. Several mechanisms have been suggested by
which either or both of these transitions may take place. There is
increasing evidence \citep[e.g.][]{wby06,wkb09,bay08,bpy08} that
\textit{strangulation}, the stripping of a diffuse gas reservoir
surrounding the galaxy \citep{ltc80}, is the most important of these
processes. Strangulation is effective already in the cluster
outskirts, and removes the gaseous envelope within a few Gyr
\citep{bcs02}. With no diffuse gas available to replenish the cold
disk gas, star formation is shut-off on similarly long time-scales.
If the density of the intra-cluster medium is high enough, also the
cold disk gas may be stripped \citep[\textit{ram-pressure stripping,
}][]{gug72,fas80,qmb00}, thus truncating star formation on a very short
time-scale. A few galaxies have been identified where ram-pressure
stripping of the star-forming gas is evident through tails of
H{\sc i} or H$\alpha$ emission and UV-bright knots
\citep{gbm01,kgv04,suv05,sdv07,cmr07}; all these galaxies are observed at
the cores of massive clusters. At lower densities, i.e. in the cluster
outskirts, ram-pressure stripping seems to be effective at truncating
the cold gas disks as traced by HI beyond the stellar disk, but
without apparent effects on the optical appearance of these galaxies
\citep{cgb90,cgk07}.

By themselves, neither strangulation nor ram-pressure stripping can
alter the morphology of infalling (spiral) galaxies, apart from the
fading of star-forming regions and spiral arms.  While galaxy mergers
are an efficient mechanism to transform spiral galaxies into
ellipticals \citep[e.g.][]{too77,fas82,new83,bah92,bah96,nab03,cdm06}, they
are unlikely to occur frequently in galaxy clusters due to the high
relative velocities of cluster members (an exception is merging with
the central galaxy). However, encounters between galaxies which do not
lead to merging can still affect galaxy properties via tidal forces
which heat the stellar disk
\cite[\textit{harassment}][]{mlk98}. Furthermore, galaxies experience
tidal forces in the gravitational potential of the
cluster, which can similarly heat the stellar disk \citep{gne03}.

These processes can be distinguished by the different effects they
have on the properties of galaxies, and by the timescale on which
galaxies are affected. For example, strangulation causes an
approximately exponential decline of the star formation rate,
ram-pressure stripping truncates star formation abruptly, whereas
mergers and tidal interaction cause gas to be transported to the
center, thus triggering bursts of enhanced star formation. Examining
the recent star formation histories of cluster galaxies is therefore a
powerful tool to identify the processes at work.

\subsection{Radial cluster profiles as density probes}

In galaxy clusters, the star formation--density relation translates to
a star formation--radius relation, since the density decreases with
increasing (projected) clustercentric radius.  Measurements of local
density, such as the distance to the $n^{\rm th}$ neighbor have the
disadvantage of strong shot noise, as well as selection and projection
effects. In this sense, radial distance presents a much cleaner
measurement. Stacking clusters allows for much better statistics than
possible with any individual cluster, and yields the properties of an
\textit{average} cluster.  Although the properties of individual
clusters can differ significantly from the average cluster properties,
even for clusters with similar masses \citep{met07}, this scatter
seems to be mostly stochastic in nature: several studies find that
there is no systematic trend of varying galaxy populations with
cluster mass, at least not for massive clusters
\citep{tgo04,bbn04,got05,fbz08}. There are, however, indications of a
dependence on cluster mass at lower masses \citep[$\lesssim 10^{13.5}
M_{\odot}$,][]{hsw08}, as well as at higher redshift \citep{pll06}.

Clustercentric distance has a second interpretation apart from a local
density measure: it also relates to the time since infall into the
cluster \citep{gwj04}. Of the galaxies at or beyond the virial radius,
a considerable fraction has not (yet) experienced the dense cluster
environment. Galaxies at the cluster core, on the other hand, were
either born in a dense environment, or they must have traversed the
cluster from the outskirts to the center at least once. The cluster
crossing time is of the order of $R_{200}/\sigma_v$, i.e. 2.5~Gyr, and
is independent of cluster mass, since $R_{200} \propto
\sigma_v$. Therefore, clustercentric distance is also an approximate
timescale, and is sensitive to processes occurring on timescales of
the order of Gyr. This timescale is particularly interesting for the
processes which have been proposed to quench star formation in
galaxies entering the cluster environment. Processes which quench star
formation on long timescales, such as strangulation or harassment,
would induce gradients over the complete radial range. Processes which
act on shorter timescales are likely to cause distinctive signatures
at the radii where they are most effective - e.g. ram-pressure stripping
could be detectable as an increase in the (post-)starburst rate,
presumably mostly in the cluster center, where the gas densities are
highest. Mergers could induce (post-)starbursts primarily at outer
cluster radii, where the relative galaxy velocities allow frequent
mergers. (Note that the truncation of star formation induces the same
spectral signatures as a true post-starburst galaxy, see
Sect.~\ref{sect:sfh}.)

The dependence of the galaxy population mix on distance from the
cluster center has been investigated in numerous studies, with an
overall consensus that the fraction of star-forming galaxies declines
towards the cluster center, as traced both by galaxy color
\citep[e.g.][]{kob01,qbb06,blb07,hsw08} and emission-line strength
\citep[e.g.][]{lbp02,gnm03,bem04,chz05}.  These studies furthermore
find that the suppression of star-formation can be traced to $\sim2-4
R_{200}$.  However, as we have motivated above, the properties of
star-forming cluster galaxies can provide additional information that
goes beyond a simple quantification of the galaxy bimodality in
clusters.
Studies of e.g. the occurrence of post-starburst galaxies have largely
been limited to intermediate-redshift clusters, with the notable
exception of \citet{hmb06}.  \citeauthor{hmb06} showed that in the
local universe, post-starburst galaxies are found in the same
environments as star-forming galaxies \citep[see also][]{zzl96}. But
they also find a marginal increase in the ratio of post-starburst to
star-forming galaxies within the virial radius of clusters.

At intermediate redshifts, the reported results are
contradictory. \citet{bmy97,bmy99} found no enhancement of starburst
or post-starburst galaxies in a sample of 15 X-ray-selected clusters
at $z\sim0.3$ . But in a sample of 10 clusters at slightly higher
redshifts, \citet{dsp99} and \citet{psd99} find a marked increase in
the occurrence of post-starbursts. Subsequent studies have not been able
to resolve these discrepant findings: evidence has been presented for
cluster-triggered post-starbursts \citep{tfi04,paz09} but also for
suppression of post-starburst in groups \citep{ynf08}. \citet{paz09}
show that post-starbursts are found both in rich clusters as well as a
in a subset of less massive groups.

\subsection{The occurrence of AGN as function of environment}

A different viewing angle to the galaxy population mix is the
occurrence of Active Galactic Nuclei (AGN). The details of the link
between AGN activity and the properties of the host galaxy,
particularly its star formation history, are still a matter of
debate. But the mere existence of the linear correlation between the
mass of the central black hole and the mass of the bulge
\citep[e.g.][]{mtr98} is a strong indication that the formation of the
two must be closely linked \citep[e.g.][]{chr99,hkb04}. Powerful
optical AGN trace current star formation activity, as they are found
predominantly in galaxies with massive, but young, bulges
\citep{kht03}. At least in the local universe, this AGN population
correlates well with X-ray selected AGN \citep{hph05}, which are thus
also likely linked to ongoing star formation. Weak optical AGN and
radio-selected AGN, on the other hand, are found predominantly in
massive galaxies with old stellar populations \citep{kht03,bkh05b}.

A prerequisite for AGN activity is a mechanism to transport gas to the
central black hole. In disk galaxies, this may take place via major or
minor mergers, or disk instabilities. Increased AGN activity in
cluster galaxies could thus trace the effect of mergers, but possibly
also harassment, which also calls for disturbance of the
disk.

However, the influence of environment on AGN activity is not yet well
established. Large-scale statistical studies of optical AGN have
become possible only with the SDSS, but a unanimous picture has not
emerged.  \citet{kwh04} showed that galaxies in dense environments are
less likely to host a powerful optical AGN ($L\mbox{[O{\sc iii}]}>10^7
L_{\odot}$), similar to the suppression of star formation, but find no
dependence on environment for weaker optical AGN. A previous study by
\citet{mng03} found no dependence of optical AGN occurrence on
environment, presumably because their sample was dominated by weak
AGN. On the other hand, \citet{pob06} report a lower fraction of (weak
and strong) optical AGN in clusters with velocity dispersions greater
than 500~km/s, compared to the field and smaller systems.

In \citet{blk07}, we investigated the occurrence of radio-selected and
optical AGN in Brightest Cluster Galaxies (BCGs) compared to non-BCGs
of the same stellar mass. We showed that BCGs are more likely to host
a radio-loud AGN, but less likely to host a (powerful) optical AGN. We
furthermore showed that enhanced radio-loudness occurs also for
non-BCGs at the cluster core, whereas optical AGN emission is
suppressed over most of the radial range to the virial radius.  On the
other hand, using a galaxy group sample extending to much lower
masses, \citet{pbm09} find that both radio and optical AGN activity
are enhanced in central galaxies vs. satellites of equal stellar mass,
and that the occurrence of optical AGN is independent of
clustercentric radius.

In the local universe, X-ray selected AGNs are rare, and there is
little evidence that the fraction of galaxies hosting an X-ray AGN
differs systematically from clusters to the field \citep{mmk07}. There
are however suggestions that the AGN fraction is higher in groups and
poor clusters compared to richer clusters \citep{gga07,smz08}. At
higher redshifts, there is evidence for a marked increase in the
fraction of cluster galaxies which host an X-ray AGN
\citep{ems07,msm09}, but also that the AGN fraction is comparable between
groups and the field \citep{ggn08}.

\subsection{This work}

In this work, we investigate the composition of the cluster galaxy
population as a function of clustercentric distance in a sample of 521
clusters at $z<0.1$ from the SDSS. Sect.~\ref{sect:data} presents the
cluster and galaxy selection, as well as our diagnostics of the recent
star formation history.  We use spectroscopic indicators of current
and past star formation rate, rather than just broad-band colors.  We
apply new high signal-to-noise spectral indicators for light-weighted
age and excess Balmer line strength (to identify post-starbursts)
developed by \citet{wkh07}. Using a diagram of light-weighted age
vs. stellar mass, we divide the galaxies into stellar mass-limited
subsamples of red (quiescent), green (transition-stage), young, and
very young galaxies.  In Sect.~\ref{sect:profiles}, we present the
dependence of several galaxy properties, including nuclear activity,
as function of clustercentric radius, and interpret these dependencies
in light of the different galaxy transformation mechanisms. We
summarize our results and conclusions in Sect.~\ref{sect:conclusion}.

Unless otherwise noted, we assume a concordance cosmology with
$\Omega_{\rm m} = 0.3$, $\Omega_{\Lambda} = 0.7$ and $H_0 = 100 \,h\,
\mbox{km/s/Mpc}$, where $h=0.7$.

\section{Data}
\label{sect:data}

\subsection{Cluster sample}
\label{sect:sample}

Here we analyze 521 clusters from the sample defined in
\citet{lbk07}. This cluster sample is derived from the C4 cluster
catalog by \citet{mnr05}, but special attention was paid in selecting
the Brightest Cluster Galaxy (BCG) as the galaxy to be most likely to
be at the center of the cluster potential well. In \citet{lbk07} and
\citet{blk07} we have shown that these galaxies indeed differ
systematically from other galaxies. The differences can well be
attributed to their special position in the cluster center, and the
rich merger history expected for central cluster galaxies. In the
absence of X-ray centroids, the position of the BCG has been shown to
be a good marker for the center, with a typical offset from the X-ray
centroid of $<$~100~kpc \citep{kma07}. We are therefore confident that
the thus selected BCGs are the best available indicators of the cluster
center for these systems. A further deviation from the original C4
catalog was the recalculation of the cluster velocity dispersions.

\begin{figure}
\begin{center}
\includegraphics[width=0.9\hsize]{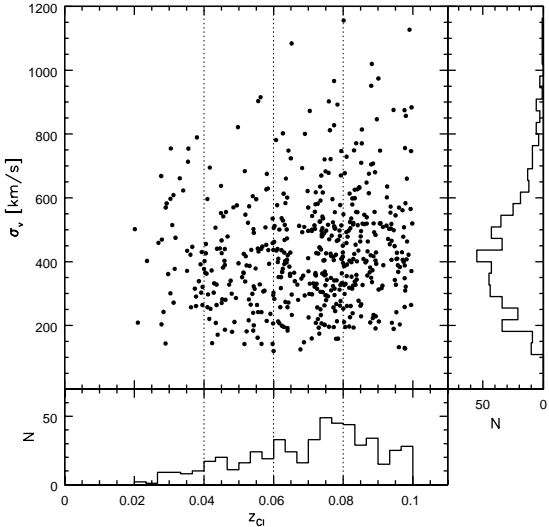}
\caption{~Distribution of the redshifts and velocity dispersions of
  the 521 clusters in our sample. Error bars have been omitted for
  clarity. The histograms show the distribution binned in redshift
  (lower panel) and in velocity dispersion (right panel). The dotted
  lines indicate the divisons for the four redshift subsamples.}
\label{fig:redshift-sigma}
\end{center}
\end{figure}

For the present work, we have further refined the sample from
\citet{lbk07} to account for the available data on cluster galaxies
and the structure of the cluster. For each cluster, we have determined
the SDSS survey coverage from the photometric catalogs \citep{vdl07},
and retain only clusters for which more than 95\% of the area within
$R_{200}$ is in the survey area, i.e. is not subject to masks from
bright stars, field edges, etc. Furthermore, for each cluster we have
determined a light-weighted centroid from galaxies in the photometric
catalog, after doing a statistical background subtraction to estimate
the likelihood of cluster membership \citep{vdl07}. In the following,
we consider only clusters for which the distance between the light
centroid and the BCG is less than 400~kpc, and less than
$0.3\;R_{200}$ . This constraint is placed to ensure that the BCG
position is a good tracer of the cluster center, and that therefore
the association between distance from the BCG and local density / time
since cluster infall rate is valid. Clusters for which this is not the
case, i.e. clusters with highly unusual structure, such as double
clusters, are excluded from the analysis. These two constraints reduce
the sample from 625 to 585 and then to 521 clusters. The redshifts and
velocity dispersions of these 521 clusters are shown in
Fig.~\ref{fig:redshift-sigma}.

\subsection{Galaxy selection}
\label{sect:geese}

For the 521 clusters in our sample, we select the galaxies within $\pm
3 \sigma_v$ and $20 R_{200}$, using the SDSS spectroscopic
database. We exclude the BCGs from the galaxy sample, as we have
shown that they differ systematically from other galaxies
\citep{lbk07,blk07}. The large radial range allows to probe
radial dependences to distances (far) beyond the turn-around
radius. If a galaxy could be associated with more than one cluster, we
assign it to the ``closest'' cluster, according to the distance
measure $\Delta=\sqrt{(d/R_{200})^2+(\Delta_v/\sigma_v)^2}$, where $d$
is the radial distance from the BCG, and $\Delta_v$ the velocity
offset.  For this association of galaxies with clusters, not only the
final 524 clusters are taken into account, but also those in the
original C4 catalog. For those clusters that were removed in the BCG
selection process, we use the velocity dispersion as given in the
original catalog. According to \citet{mnr05}, the C4 cluster catalog
is about 90\% complete. Hence, our sample of field galaxies should be
quite clean from galaxies in undetected clusters. Since the C4 catalog
was compiled from DR3, only galaxies contained in DR3 are used for the
field sample. Within $R_{200}$, these data are supplemented with DR4
data, for which the MPA/JHU value added catalogues\footnote{\tt
  http://www.mpa-garching.mpg.de/SDSS/} were available at the time of
writing.

\subsection{Spectral indices}
\label{sect:sfh}

\begin{figure*}
\begin{center}
\includegraphics[trim=1cm 0 0.5cm 0,width=0.47\hsize]{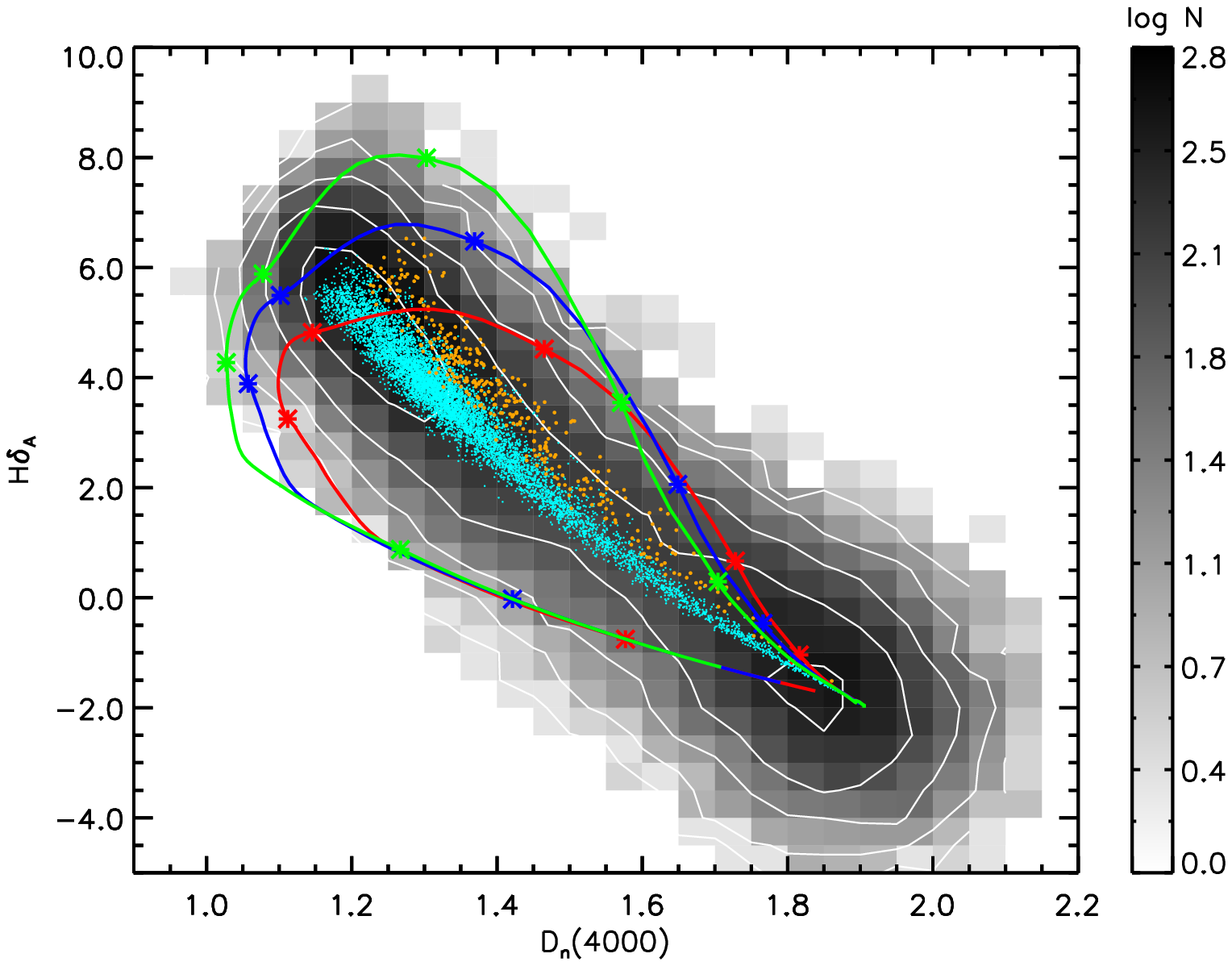}
\hspace{0.03\hsize}
\includegraphics[trim=1cm 0 0.5cm 0,width=0.47\hsize]{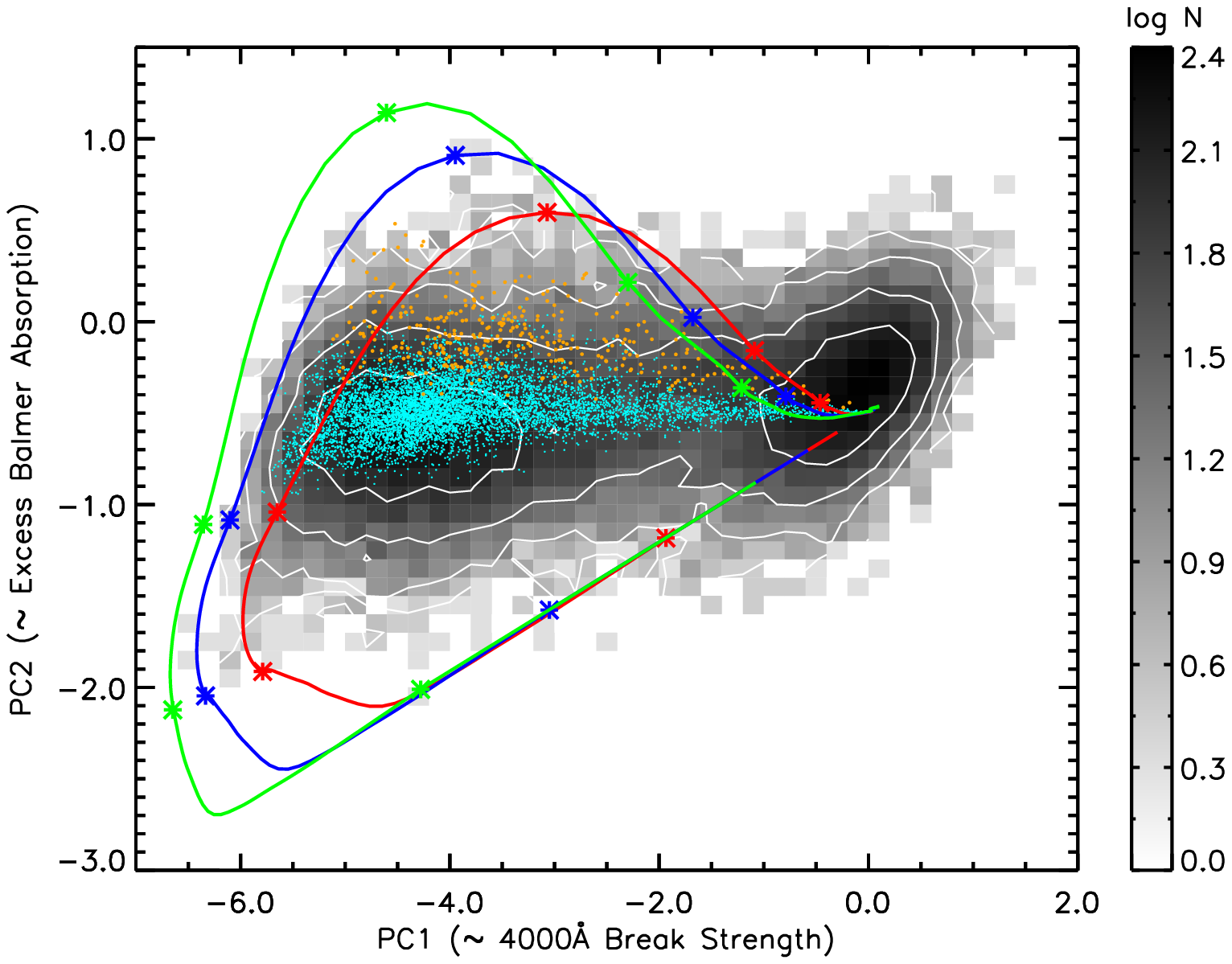}
\caption{In grayscale, the joint distributions of H$\delta$ vs.
  D$_n$(4000) (left) and PC2 vs. PC1 (right) for galaxies in our
  sample with masses $<10^{10.5}$ and redshifts $>0.06$.  The red
  sequence lies on the right of each diagram, and the blue sequence
  extends to lower D$_n$(4000) or PC1. Starburst galaxies are found at
  the bottom left of each diagram, with weak Balmer absorption and
  weak 4000\AA\ break strengths. Post-starburst and truncated star
  formation galaxies lie above the blue sequence, with stronger Balmer
  absorption than expected for their 4000\AA\ break strength. To aid
  in interpretation of the figure, we have created examples of
  population synthesis models with different star formation history
  scenarios. Overplotted as cyan dots are model galaxies composed of
  an old bulge population, and a young disk population with disk:bulge
  ratios varying from 0 to 1. The star formation history of the disks
  fluctuates around a constant, creating the scatter observed in the
  distribution. The purple points show models in which star formation
  has decreased substantially over a short timespan. This causes the
  models to move upwards in PC2 and diagonally to the top right in
  D$_n$(4000)/H$\delta$. Overplotted as tracks are bulge+starburst
  models with burst mass fractions (defined as the total mass created
  in the burst divided by the total mass in the bulge) of 5\% (red),
  10\% (blue) and 20\% (green). Stars indicate time intervals 0.001,
  0.01, 0.1, 0.5, 1.0, and 1.5 Gyr after the start of the burst (the
  tracks run clockwise from the red sequence in both diagrams).}
\label{fig:tracks_pc1_pc2_d4_hd}
\end{center}
\end{figure*}

Our main focus in this analysis is the recent star formation history
of cluster galaxies. Different spectral features are associated with
different timescales of star formation. In this paper we make use of
nebular emission lines which indicate star formation rates on
timescales of $10^7$years, the Balmer absorption lines for timescales
of $\sim0.5$\,Gyr and the 4000\AA\ break strength for longer
timescales of a few Gyr. For the latter two indicators, we make use of
the spectral indices developed by \citet{wkh07}, which are based on a
Principal Component Analysis of the wavelength region
3750-4150\AA. The first principal component (PC1) is essentially
equivalent to the strength of the 4000\AA$\,$ break, and is thus a
measure of the light-weighted mean age of the stellar population. The
second principal component (PC2) measures the \textit{excess} Balmer
absorption compared to the expectation value for a given 4000\AA\,
break strength (the locus of galaxies with continuous star
formation). Traditionally, the Balmer absorption line strength is
quantified as the strength of the H$\delta$ line - the great advantage
of the new index is that it can be applied to spectra with
considerably lower SNR than required for measuring H$\delta$
\citep{wkh07}.

Fig.~\ref{fig:tracks_pc1_pc2_d4_hd} illustrates the distribution of a
subsample of the SDSS galaxies used in this paper, in PC2 vs. PC1 and
the more traditional H$\delta$ vs. D$_n$(4000) for comparison. On the
right hand side of both figures, the red sequence can be seen with
large 4000\AA\ break strengths (PC1$\sim0$) from the predominantly old
stellar populations. As we move left in both diagrams, the mean
light-weighted stellar age decreases, indicated by a decrease in
4000\AA\ break strength, and in this dataset a well defined blue
sequence is seen at PC1$\sim-4$. To the bottom left of both diagrams
we find weak Balmer lines and weak break strengths i.e. starburst
galaxies. And above the blue sequence, with strong Balmer absorption
lines, we find the ``Balmer strong'' galaxies, which are often
associated with post-starburst galaxies, or galaxies which have
suffered a recent truncation in their star formation history.  Because
PCA identifies lines of correlation in the dataset, and H$\delta$ and
D$_n$(4000) are correlated in galaxy populations, the first component
of PCA naturally takes account of this correlation. This causes the
twist of the figures relative to each other, and means that PC2 is
equivalent to ``excess'' Balmer absorption over that expected for the
4000\AA\ break strength of the galaxy. Along with the improved SNR of
the PCA indicators, this makes the identification of Balmer-strong
galaxies easy, as they are displaced upwards in PC2 from the main
cloud of points.

To aid interpretation of both diagrams we have overplotted examples of
stellar population synthesis models (from Charlot \& Bruzual 2009, in
prep.), with differing star formation scenarios.  The cyan points
indicate the position of galaxies composed of the superposition of two
components with differing star formation histories: an old ``bulge''
component, and a young ``disk'' component. Disk-to-bulge ratios vary
between 0 and 1, and the SFR of the disks is allowed to fluctuate on
timescales of $10^8$ years which causes the scatter in the
distribution compared to standard smoothly exponentially declining
SFHs ($\tau$ models). The purple points are a subset of the cyan
points - in these models, star formation has decreased substantially
in a short timespan, i.e. has been truncated. Note how these star
formation histories can easily be distinguished from the ``normal''
models by their larger PC2 values.
 The tracks show a different star
formation scenario: the superposition of short starbursts on old
stellar populations. Galaxies then move in a clockwise circle, from
the red sequence, quickly to the starburst phase, then climb
increasingly slowly in PC2 or H$\delta$ into the post-starburst phase,
until the burst fades and they reenter the red sequence. As we can
see, the indices PC1/PC2 are interesting for investigating the stellar
populations of cluster galaxies, since they can test whether cluster
galaxies are prone to experience starbursts (e.g. through mergers) or
truncation (by ram-pressure stripping) more often than field galaxies.

\begin{figure*}
\begin{center}
\includegraphics[width=0.48\hsize]{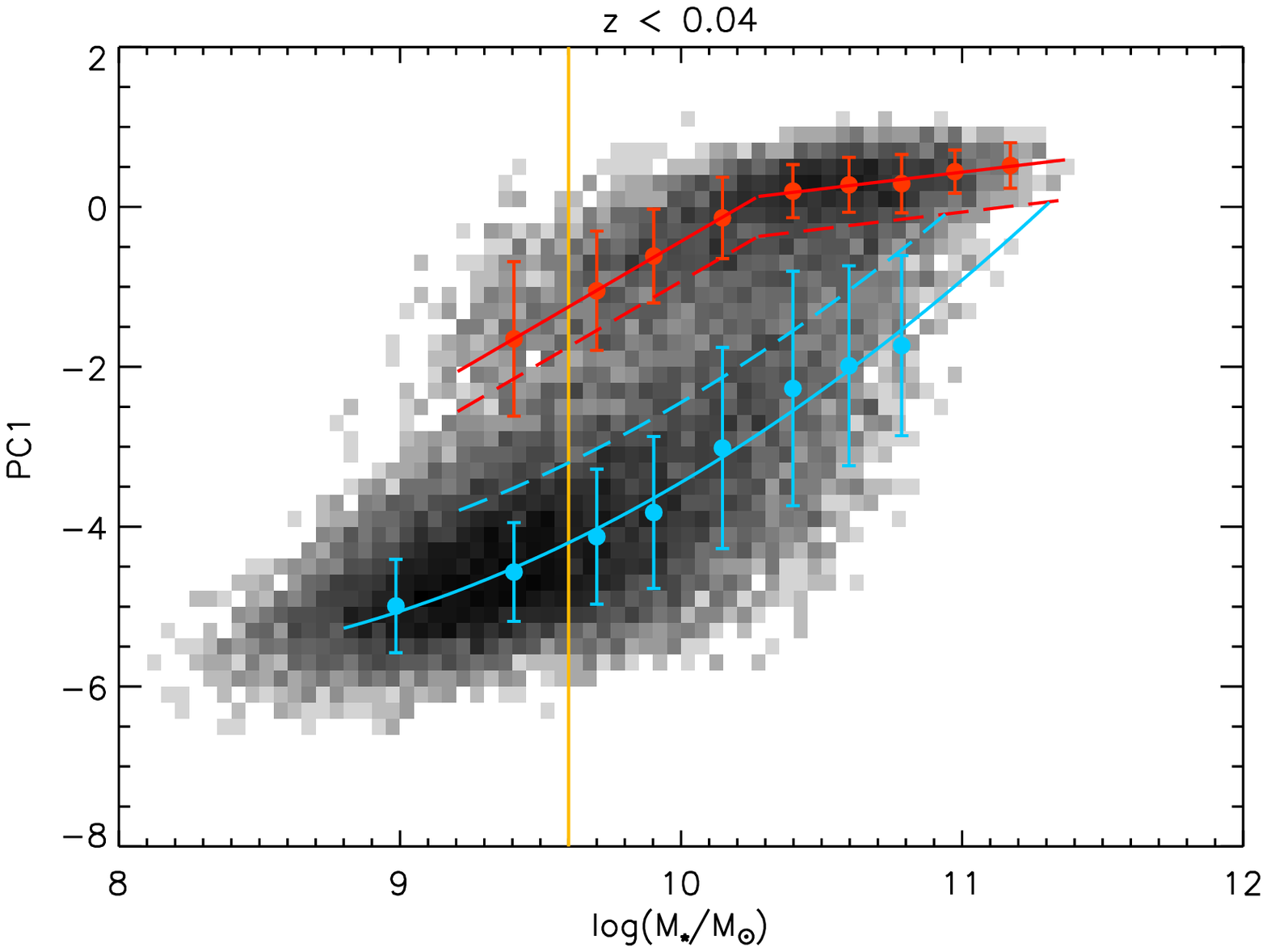}
\includegraphics[width=0.48\hsize]{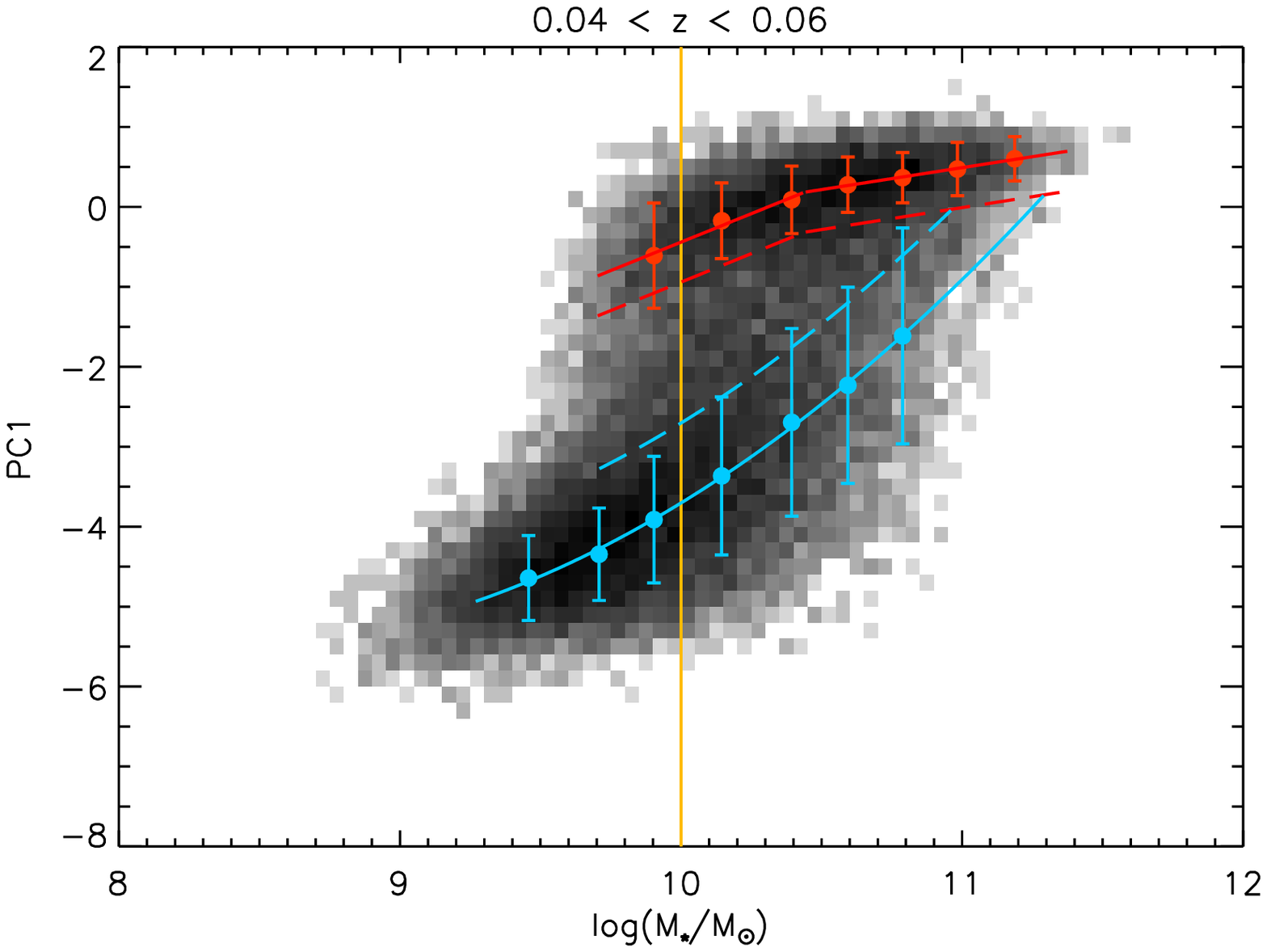}
\includegraphics[width=0.48\hsize]{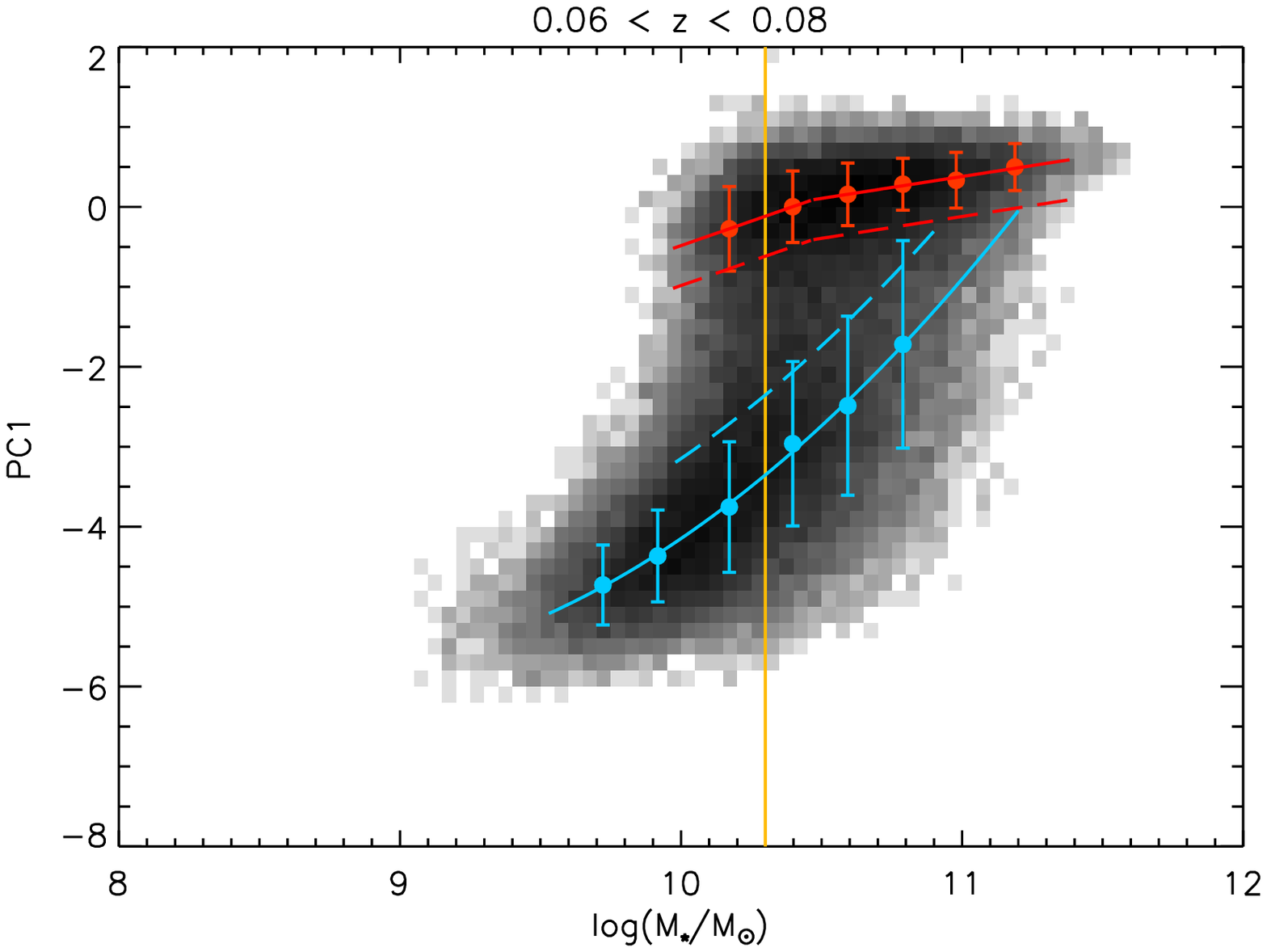}
\includegraphics[width=0.48\hsize]{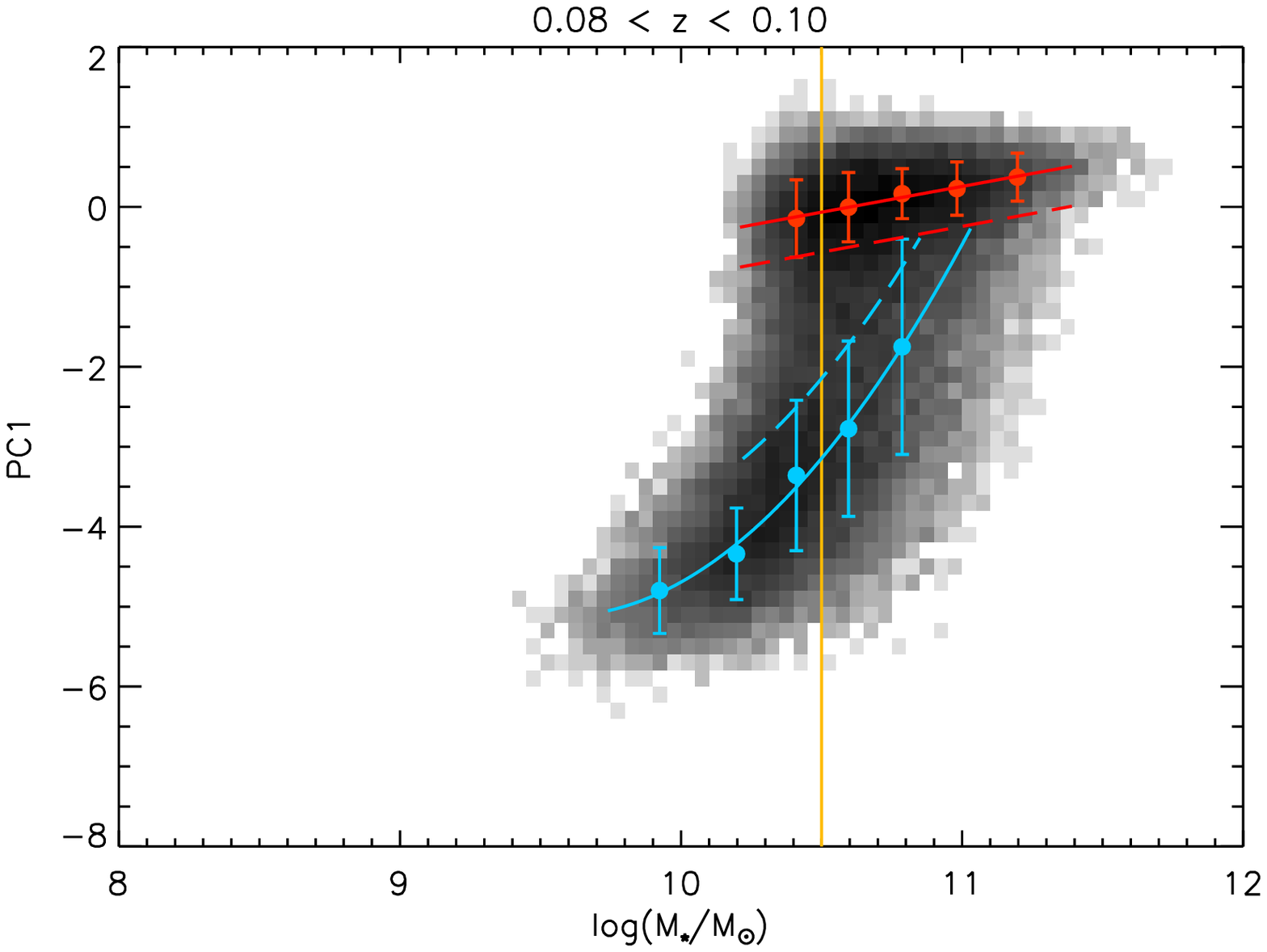}
\caption{ ~The distribution of PC1 vs. stellar mass for all galaxies
  in our sample, in the four redshift subsamples, shown as
  logarithmically scaled grayscale. For each mass bin, the red (blue)
  circle shows the peak of the Gaussian curve fitted to the
  distribution of the red (blue) sequence, and is shown at the median
  mass of the particular mass range. The error bars indicate the
  respective $1\sigma$ width. The dashed red and blue lines indicate
  the limits of the green valley: galaxies above the dashed red line
  are classified as red, between the red and blue dashed lines as
  green, between the dashed and solid blue line as cyan, and below the
  solid blue line as blue (see
  Sect.~\ref{sect:spectro:classification}) . The vertical orange line
  indicates the mass limit for red galaxies in each sample.  }
\label{fig:mass_pc1_subsamples_2}
\end{center}
\end{figure*}

\subsection{PC1 vs. stellar mass}

Fig.~\ref{fig:mass_pc1_subsamples_2} presents the distribution of the
galaxies in our sample in the PC1 vs. stellar mass diagram in four
redshift subsamples (see next section).  The stellar masses have been
determined from the {\tt model} magnitudes, via the {\tt
  kcorrect\_v4.1.4} algorithm \citep{blr07}. Both a red sequence
(strong 4000\AA$\;$ break, PC1$\sim0$) and a blue sequence
(PC1$\sim-4$) are clearly visible.

This diagram is analogous to the more traditional color-magnitude
diagrams (CMDs), in that both color and PC1 are indicators of the
recent star formation history, whereas luminosity and stellar mass
measure the ``size'' of the system.  PC1 is a better indicator of the
light-weighted age of the stellar population than broad-band colors,
because it spans a much shorter range in wavelength, and is thus much
less affected by dust extinction. Stellar mass is arguably a more
fundamental galaxy property than luminosity, as the latter is highly
dependent on the age of the stellar population: young stars are very
luminous, hence galaxies with young stellar populations have low $M/L$
ratios. Furthermore, the $M/L$ ratio varies with the observed band:
blue bands are particularly dominated by young stars, whereas red
bands are sensitive also to older stellar populations. This means
effectively that the red and blue sequences are shifted in CMDs of
different bands.

\begin{figure*}
\begin{center}
\includegraphics[width=0.41\hsize]
{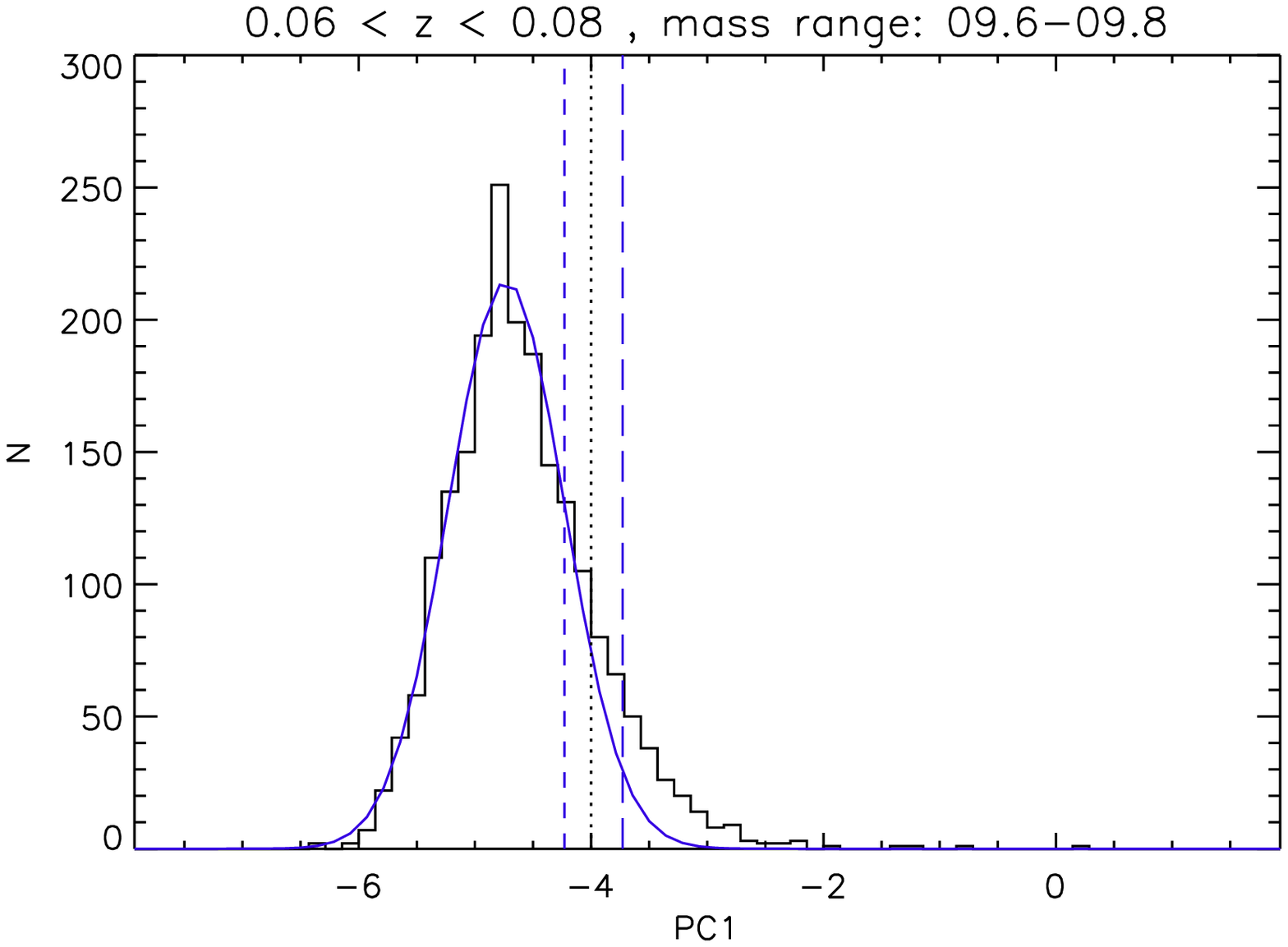}
\includegraphics[width=0.41\hsize]
{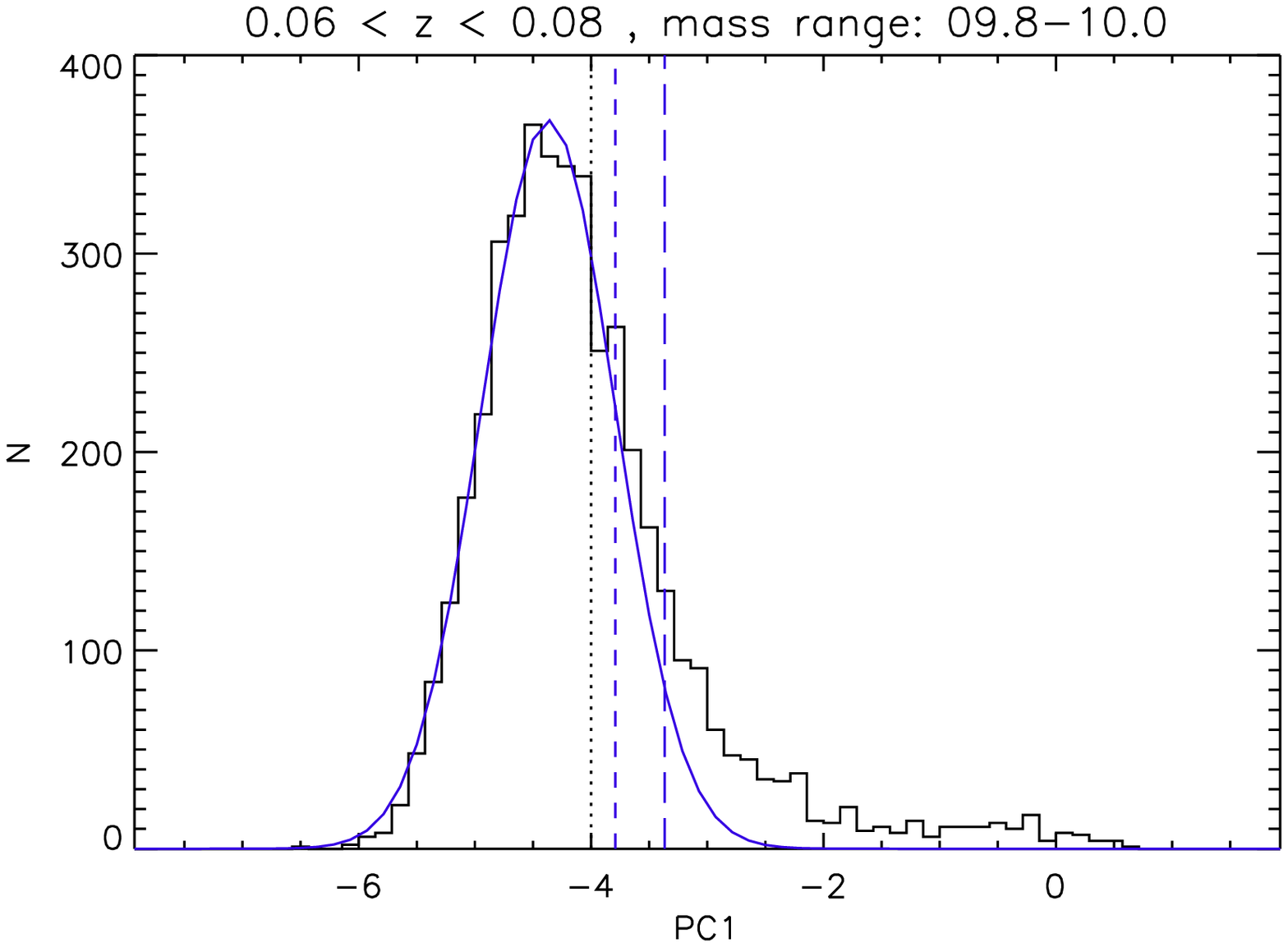}
\includegraphics[width=0.41\hsize]
{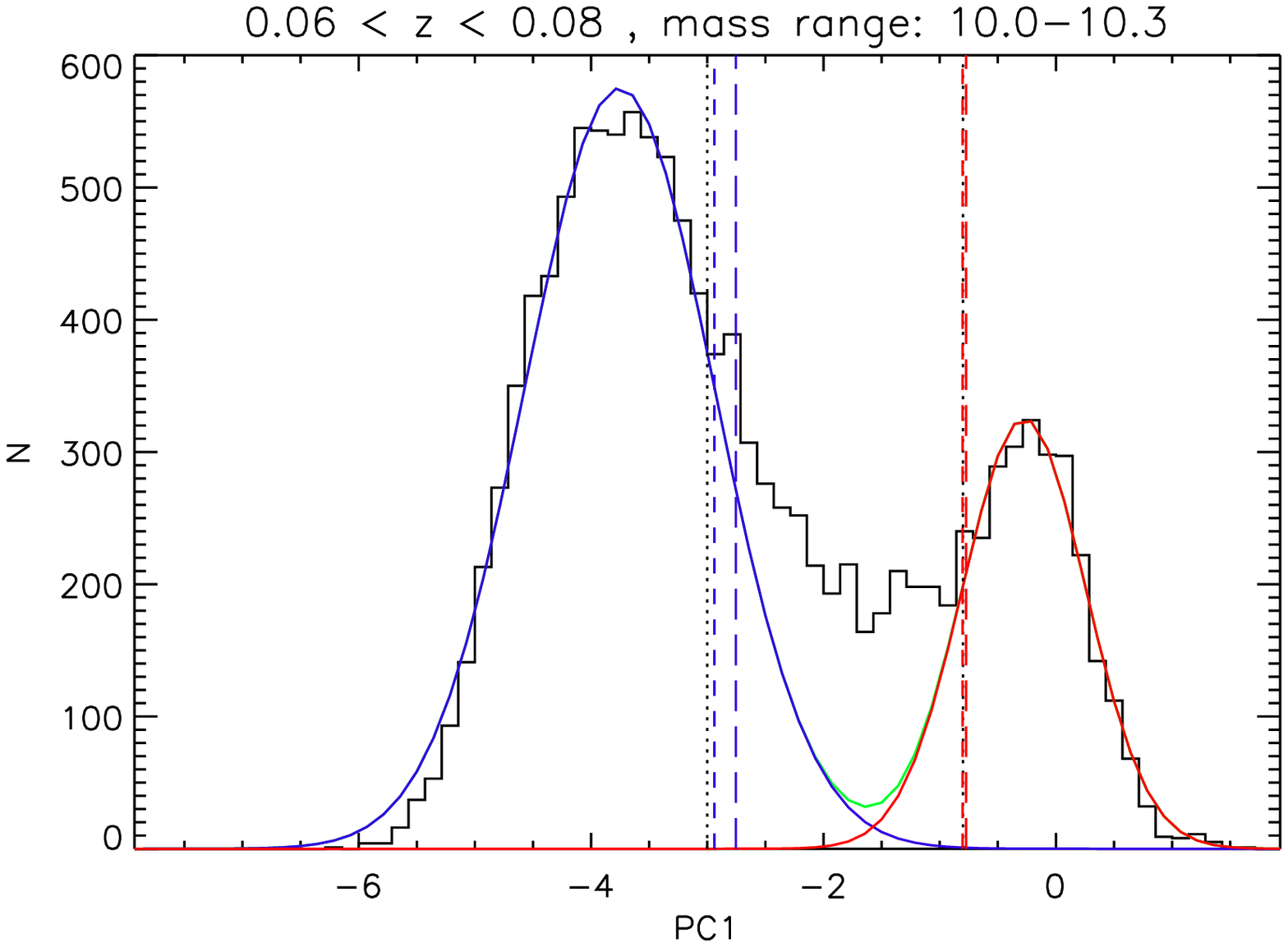}
\includegraphics[width=0.41\hsize]
{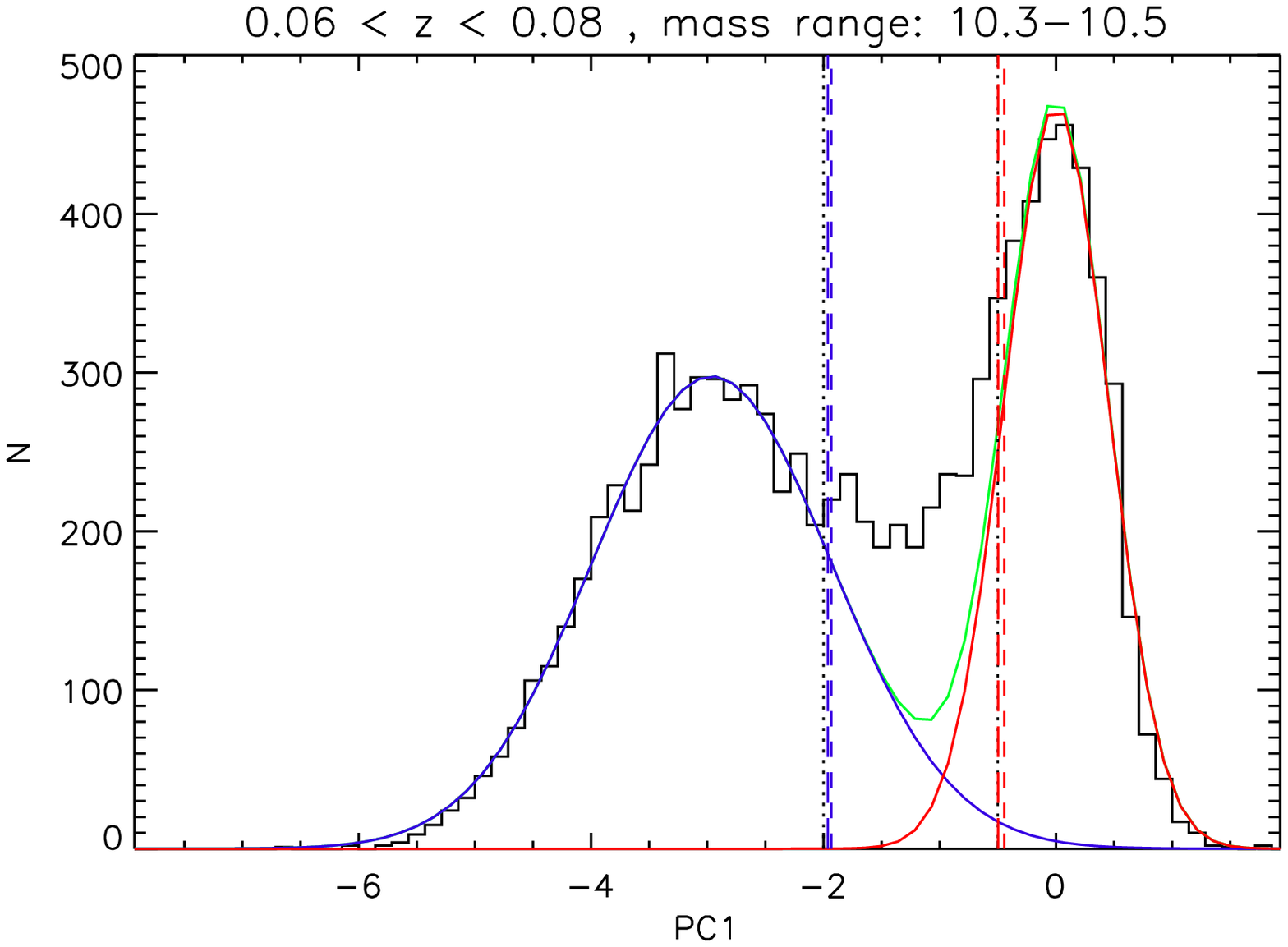}
\includegraphics[width=0.41\hsize]
{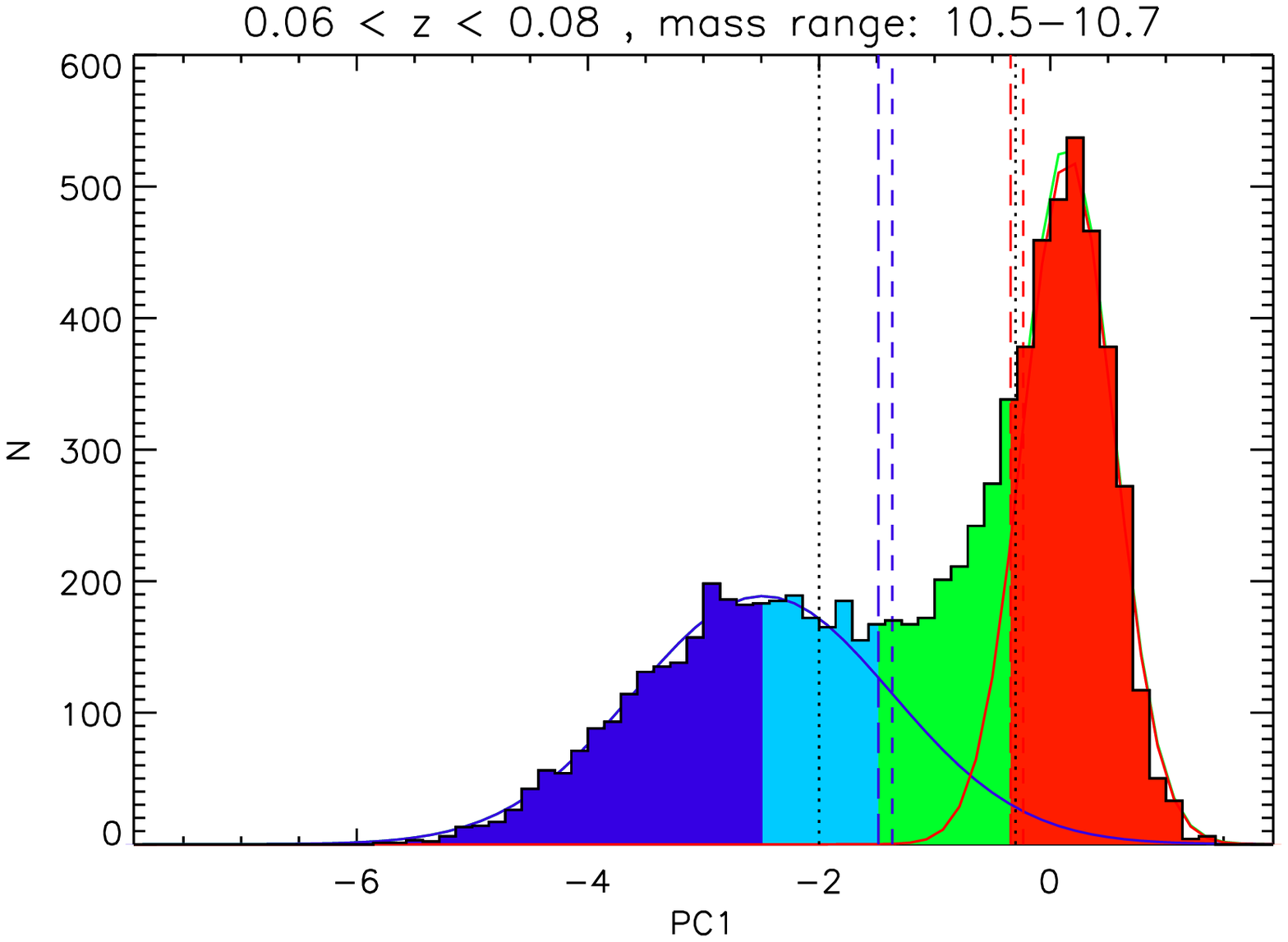}
\includegraphics[width=0.41\hsize]
{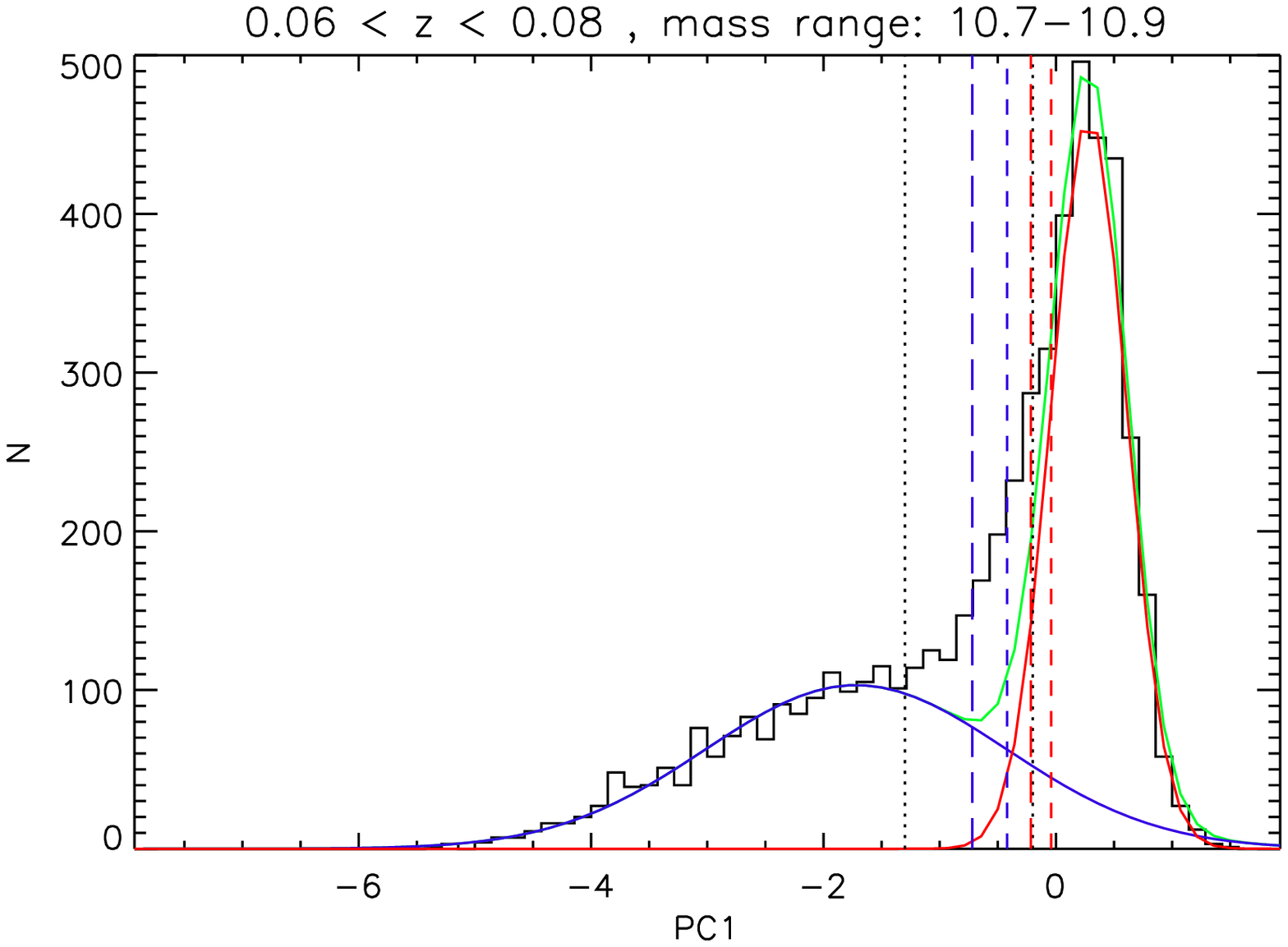}
\includegraphics[width=0.41\hsize]
{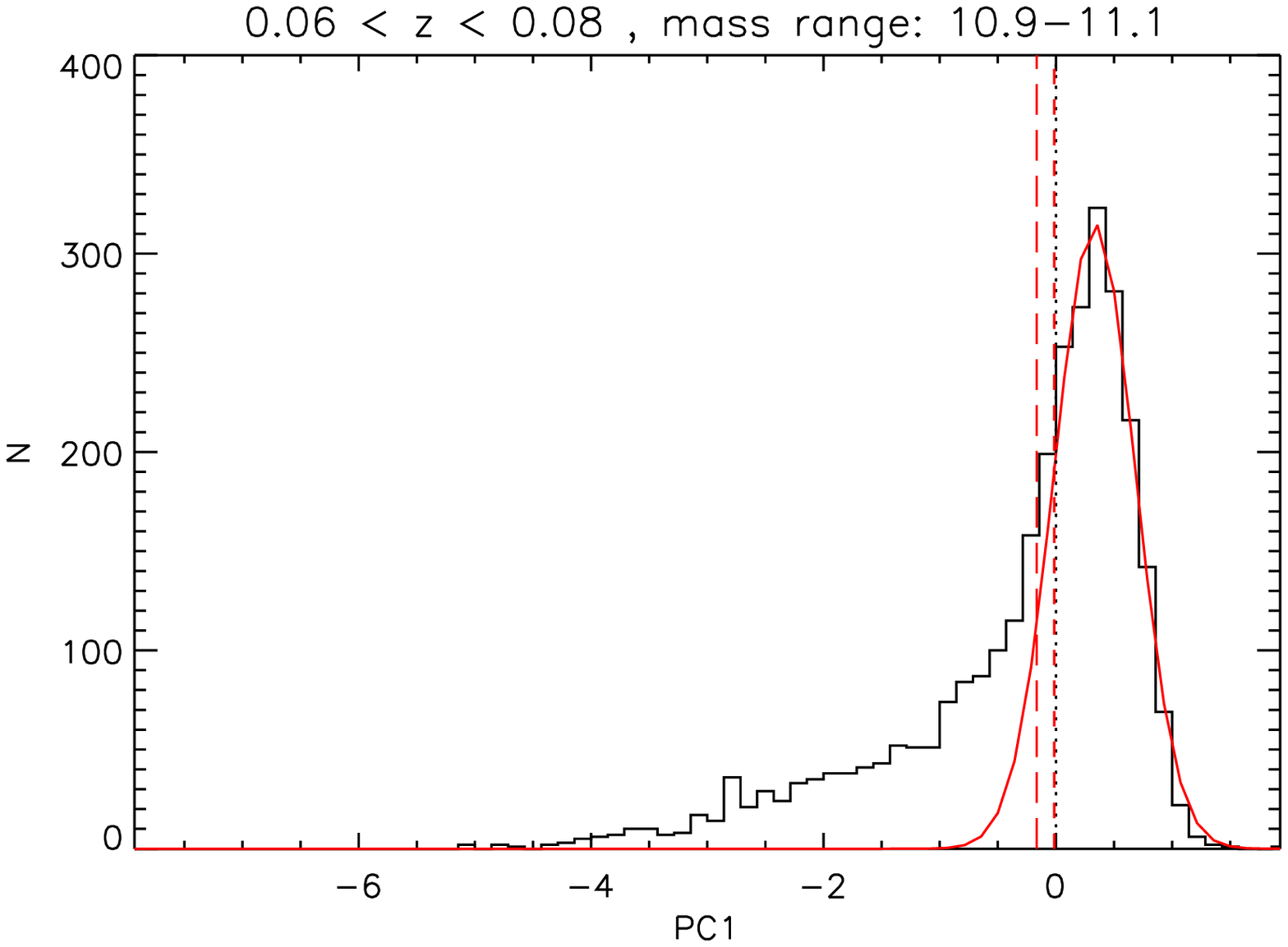}
\includegraphics[width=0.41\hsize]
{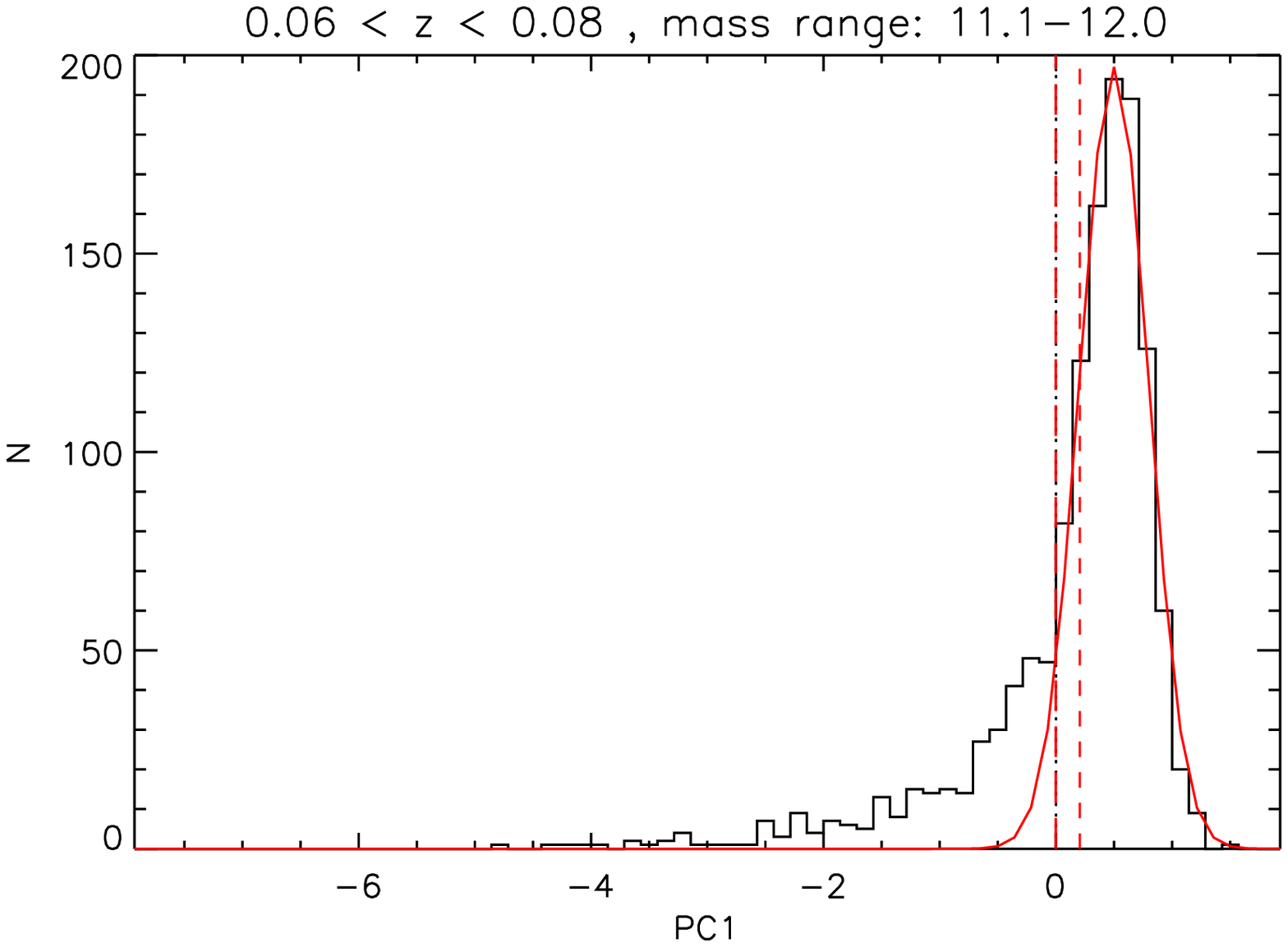}
\caption{~The distribution in PC1 for galaxies in the $0.06<z\le0.08$
  cluster sample.  The distributions are shown as black
  histograms. The green line shows the fit of a bimodal Gaussian to
  the data, where the region between the black dotted lines has been
  left out. The blue and red lines show the two individual Gaussian
  components. The short-dashed red and blue lines indicated the
  $1\sigma$ widths of the red and blue sequence (only towards the
  green valley).  The long-dashed red and blue lines show our adopted
  limits for the green valley - these are placed one PC1 above the
  peak of the blue sequence and half a PC1 unit below the peak of the
  red sequence.  As can be seen e.g. from the
  $10.7<\log(M_{\star}/M_{\odot})<10.9$ mass bin, the $1\sigma$ width
  of the blue sequence falls into the PC1 range with an excess of
  green galaxies, whereas the fixed-PC1-width cut correctly identifies
  this excess. The plot for the mass range $10.5<\log
  M_{\odot}\le10.7$ illustrates how these fits are used to classify
  the galaxies as blue, cyan, green, or red. Note that the mass ranges
  shown in the top two panels are below the mass completeness limit of
  red galaxies.}
\label{fig:gaussians_mass_pc1_0.08}
\end{center}
\end{figure*}

\citet{bgb04} fitted bimodal Gaussians to the red and blue sequences
in a color-magnitude diagram of a large sample of SDSS galaxies in
bins of luminosity. We follow the same approach here and fit the
distribution in PC1 in bins of stellar mass, shown in
Fig.~\ref{fig:gaussians_mass_pc1_0.08}.  Clearly, the distributions
are not well described by two Gaussians - there is an excess of
galaxies between the red and blue peaks. In CMDs, the region between
the red and blue sequences has been coined the \textit{green valley},
and we adopt this term here, as well.  We fit two Gaussian curves to
the distribution, but ignoring the green valley region for the
fit. The fits provide adequate descriptions for the blue edge of the
blue sequence and the red edge of the red sequence. However, in the
green valley, there is clearly a population of galaxies not accounted
for by these two components. The clear separation of the red and blue
sequence, and identification of the green valley, illustrates the
advantage of using spectroscopic age indicators over broad-band
colors, cf. Figs.~3 and 4 of \citet{bgb04}.

Fig.~\ref{fig:mass_pc1_subsamples_2} also indicates the peaks and
$1\sigma$ widths of the red and blue sequences. The ridgeline of the
red sequence can be well described by two linear relations (the
highest redshift bin does not probe low enough masses, hence only one
line is fit), whereas the blue ridge is better described by a
parabola. With these fits, we have continuous descriptions of the red
and blue sequences with mass: ${\rm PC1_{RS}} (\log
M_{\star}/M_{\odot})$ refers to the ridgeline of the red sequence,
${\rm PC1_{BS}} (\log M_{\star}/M_{\odot})$ to that of the blue
sequence.

\subsection{Galaxy classification}
\label{sect:spectro:classification}

The widths of both the red and blue sequences change with mass, most
notably for the blue sequence
(Fig.~\ref{fig:gaussians_mass_pc1_0.08}). This makes it unsatisfactory
to define the extent of each sequence in terms of the widths; e.g. in
the higher mass subsets, the green valley region is within $1-2\sigma$
of the peak of the blue sequence, simply because the blue sequence is
very broad. Instead, we use a fixed offset from each ridge to separate
the three regions of the distribution. We define the upper (red) limit
of the green valley to be half a unit in PC1 lower than the peak of
the red sequence, and the lower (blue) limit to be one unit larger
than the peak of the blue sequence. As illustrated in
Fig.~\ref{fig:mass_pc1_subsamples_2} and
Fig.~\ref{fig:gaussians_mass_pc1_0.08}, these limits correspond
closely to the $1\sigma$ width of both sequences for most of the mass
range considered here. Furthermore, these limits reliably select the
excess of green galaxies over the two gaussian fit in those mass bins
where it is apparent (Fig.~\ref{fig:gaussians_mass_pc1_0.08}). By
comparison, the $1\sigma$ width of the blue sequence fails to do so at
large stellar masses.  Each galaxy is classified via the following
scheme (also illustrated in Fig.~\ref{fig:gaussians_mass_pc1_0.08}):
\begin{enumerate}
\item If ${\rm PC1 > PC1_{RS}-0.5}$, the galaxy is classified as
  ``red''. These are the quiescent galaxies, in which star formation
  has ceased $\gtrsim 1$~Gyr ago.
\item If ${\rm PC1_{BS}+1 < PC1 < PC1_{RS}-0.5}$, the galaxy is
  classified as ``green'', i.e. in the transition regime between
  young, star-forming and old, quiescent.
\item If ${\rm PC1_{BS} < PC1 < PC1_{BS}+1}$, the galaxy is classified
  as ``cyan''. Note that we have divided the young, blue population
  into two subsets - the ``cyan'' galaxies have slightly older
  populations than the ``blue'' galaxies.
\item If ${\rm PC1 < PC1_{BS}}$, the galaxy is classified as
  ``blue''. These are the youngest galaxies.
\end{enumerate}
While ``green'' may actually signify an intermediate class of galaxies
between the red and blue sequence, the term ``cyan'' has been
introduced only to ease terminology, and ``cyan'' should be considered
a subclass of blue galaxies in the more general sense. Although PC1 is
strictly speaking a measure of the age of the stellar populations,
``blue'' galaxies generally have higher levels of star formation than
``cyan'' galaxies.

It is important to make the classification in this order: the red
sequence is well defined in all mass bins (above the mass limit), but
according to this definition, not every mass bin has a green
valley. Extreme cases of this are the highest mass bins, where it is
not possible to fit a peak to the blue sequence.

The most important classification for the purpose of this paper is the
distinction between old, quiescent (red) galaxies, and young,
star-forming galaxies. We have verified that our results do not depend
on the precise location of the cut between green and red galaxies. In
particular, all our results apply qualitatively also to the red side
of the red sequence, i.e. galaxies redder than the red peak.

\begin{figure*}
\begin{center}
\includegraphics[width=0.48\hsize]{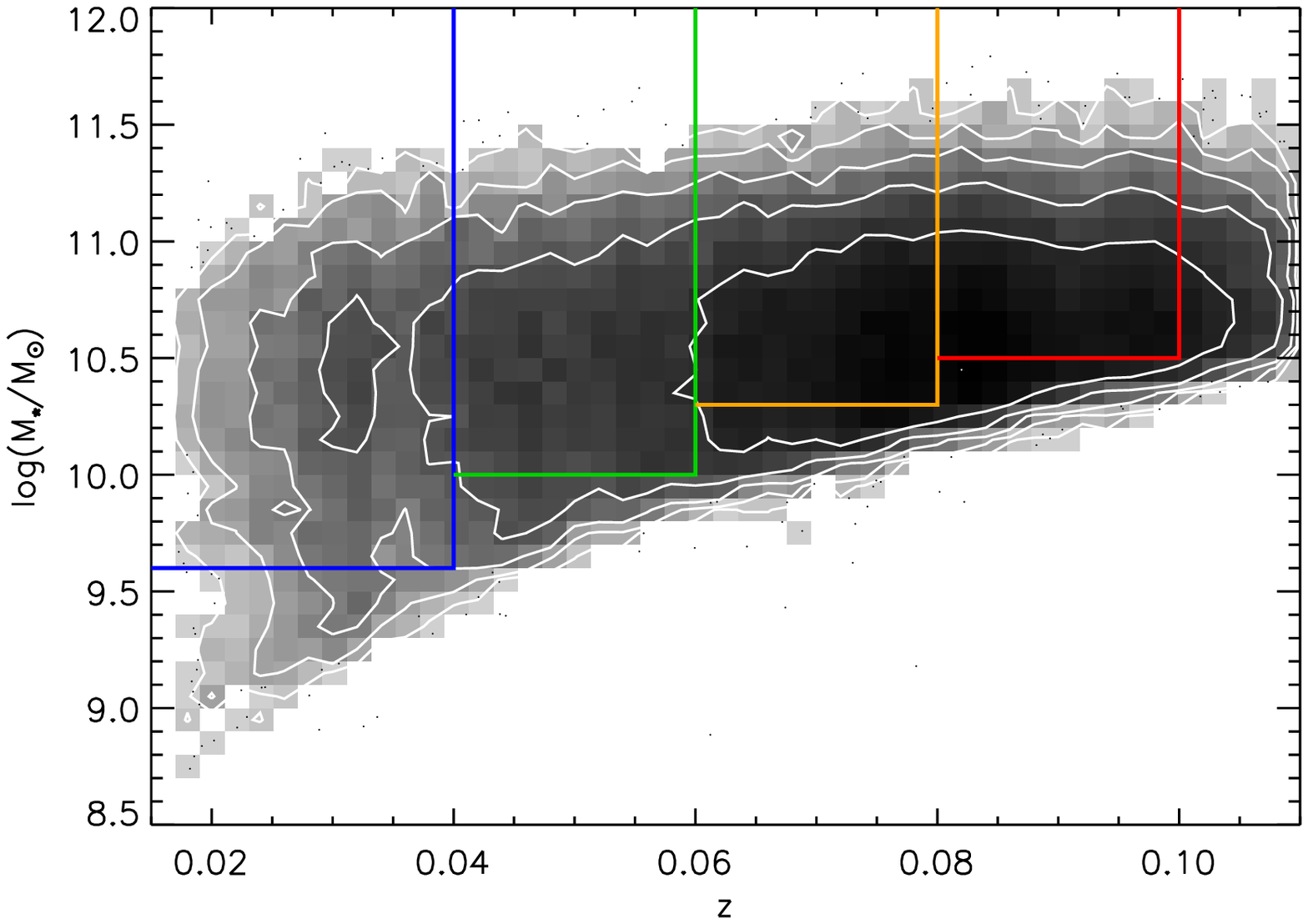}
\includegraphics[width=0.48\hsize]{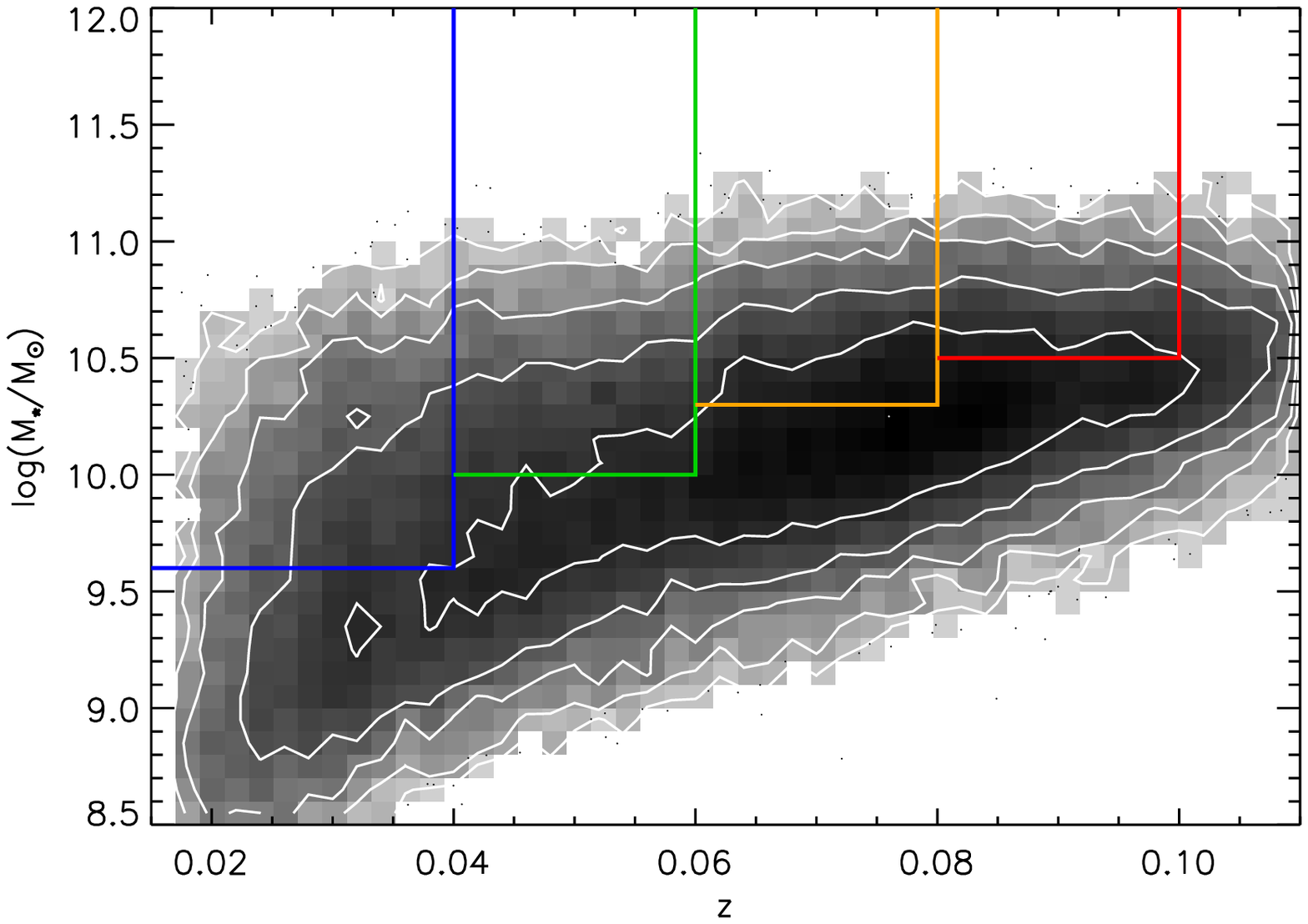}
\caption{~The stellar masses of galaxies on the PC1 red/blue sequence
  (left/right panel) against their redshift. The four boxes describe the
  four redshift subsamples described in the text, each of which is complete
  in stellar mass to the lower limit indicated by the lower boundary of each
  box.}
\label{fig:z_mass_samples}
\end{center}
\end{figure*}

\subsection{Mass completeness}
\label{sect:mass_completeness}

Our aim is to work with a sample of galaxies complete in stellar mass,
in order to follow the evolution of galaxies at fixed mass. The
galaxies with the highest mass-to-light ratios are those with old
stellar populations, i.e. the red galaxies .  In
Fig.~\ref{fig:z_mass_samples} we plot the stellar masses of galaxies
on the red sequence as a function of redshift.  For the whole sample,
the sample is complete in red galaxies only for
$\log(M_{\star}/M_{\odot})> 10.5$. But by defining subsamples of
clusters (and their galaxies) in redshift bins, the completeness limit
can be pushed to lower mass values.  We define four such samples: the
lowest redshift one ($z<0.04$) is complete in red galaxies with
stellar masses as low as $\log(M_{\star}/M_{\odot})>9.6$, but it
contains only 39 clusters. The sample with $0.04<z<0.06$ (107
clusters) includes red galaxies with
$\log(M_{\star}/M_{\odot})>10.0$. The sample in the range
$0.06<z<0.08$ (200 clusters) is complete for
$\log(M_{\star}/M_{\odot})>10.3$, and the highest redshift sample
$0.08<z<0.10$ (175 clusters) is complete only for
$\log(M_{\star}/M_{\odot})>10.5$.

Note that we refrain from using a $V_{\rm max}$ weighting. This is
because $V_{\rm max}$ assumes that galaxies are selected solely
because of their apparent magnitude. This is not the case for our
cluster galaxies, as also the parent clusters are subject to selection
biases. Furthermore, the effect of fiber collisions is not constant
with redshift - at higher redshifts, fewer galaxies in denser
environments can be targetted.

\subsection{Aperture bias}
\label{sect:spectr:aperture}

In the PC1 -- stellar mass diagram, the red and the blue sequences
are not entirely parallel. At low stellar masses, the blue sequence is
approximately parallel to the red sequence. But at masses
$\log(M_{\star}/M_{\odot}) \gtrsim 10.3$, the blue sequence curves
upward towards the red sequence. This is a consequence of the aperture
bias inherent to the SDSS survey, which obtains spectra within a
$3\arcsec$ diameter. In our lowest redshift sample, $3\arcsec$
corresponds to $\sim 2-3$kpc, and thus only the inner regions of large
(resp. massive) galaxies. Many massive disk galaxies also have a bulge
component, with a stellar population older than the one in the
disk. If only the inner part of the galaxy is observed, then the light
of the bulge with its old stellar population is dominant. This is the
reason the blue sequence turns ``upward'' with larger masses:
generally, the bulge-to-total fraction increases with stellar mass for
disk galaxies. Thus, the more massive the galaxy, the more prominent
the bulge, and the older the light-weighted population in the
bulge. Note that in galaxy samples based on slit spectroscopy, the red
and blue sequences remain parallel and well-separated over all mass
ranges \citep{vdl07}.

The aperture bias could thus ``contaminate'' our sample in cases where
the light from the central $3\arcsec$ is not a good estimator of the
total population. This would be the case e.g. for typical Sb galaxies,
with on old, red bulge, and a young, blue disk. In the three redshift
samples at $z>0.04$, 98\% of the galaxies identified as red from their
PC1 values also have a red overall color; in the $z<0.04$ sample, this
number is 90\%. Furthermore, only a small fraction ($\sim$6\%) of
PC1-red galaxies have a dominant photometric disk component
(\texttt{frac\_DeV}$ < 0.5$); of these galaxies, about 30-50\% of
these are indeed face-on, early-type spiral galaxies. The overall
``contamination'' of the red sequence galaxies is therefore rather
small, and of the order of 2-5\%.  We are thus confident that although
the aperture bias distorts the PC1--mass diagram, our classification
scheme is not significantly affected.

The aperture bias further leads to an apparent shift of the location
of the blue sequence with redshift, since the bias towards the central,
oldest bulge population is largest in the closest galaxies. This is
the reason why we need to fit the location of the blue and red
sequence in separate redshift bins.

\section{Results: radial profiles of the galaxy population}
\label{sect:profiles}

\begin{figure*}
\begin{center}
\includegraphics[width=0.48\hsize]{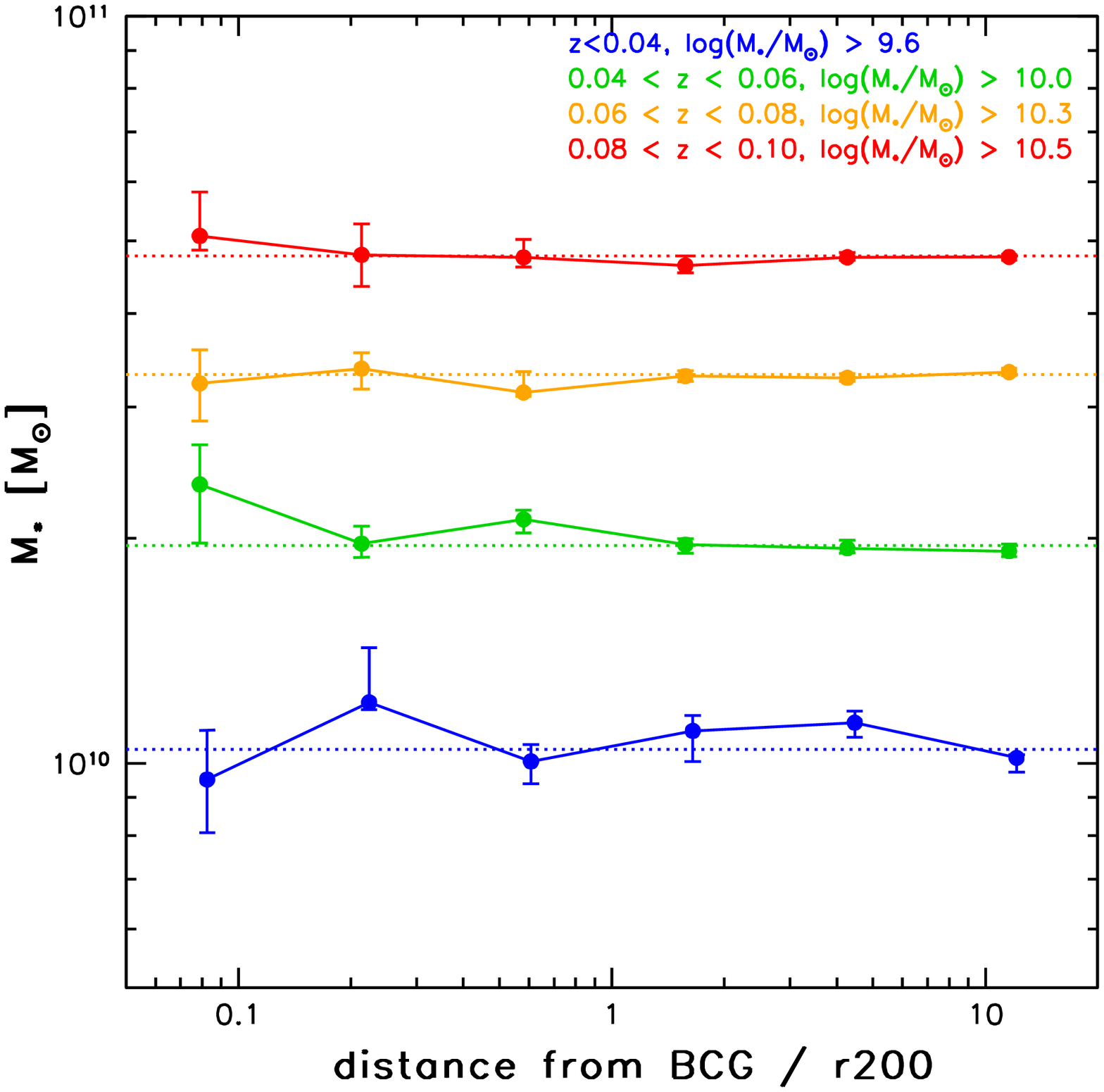}
\includegraphics[width=0.48\hsize]{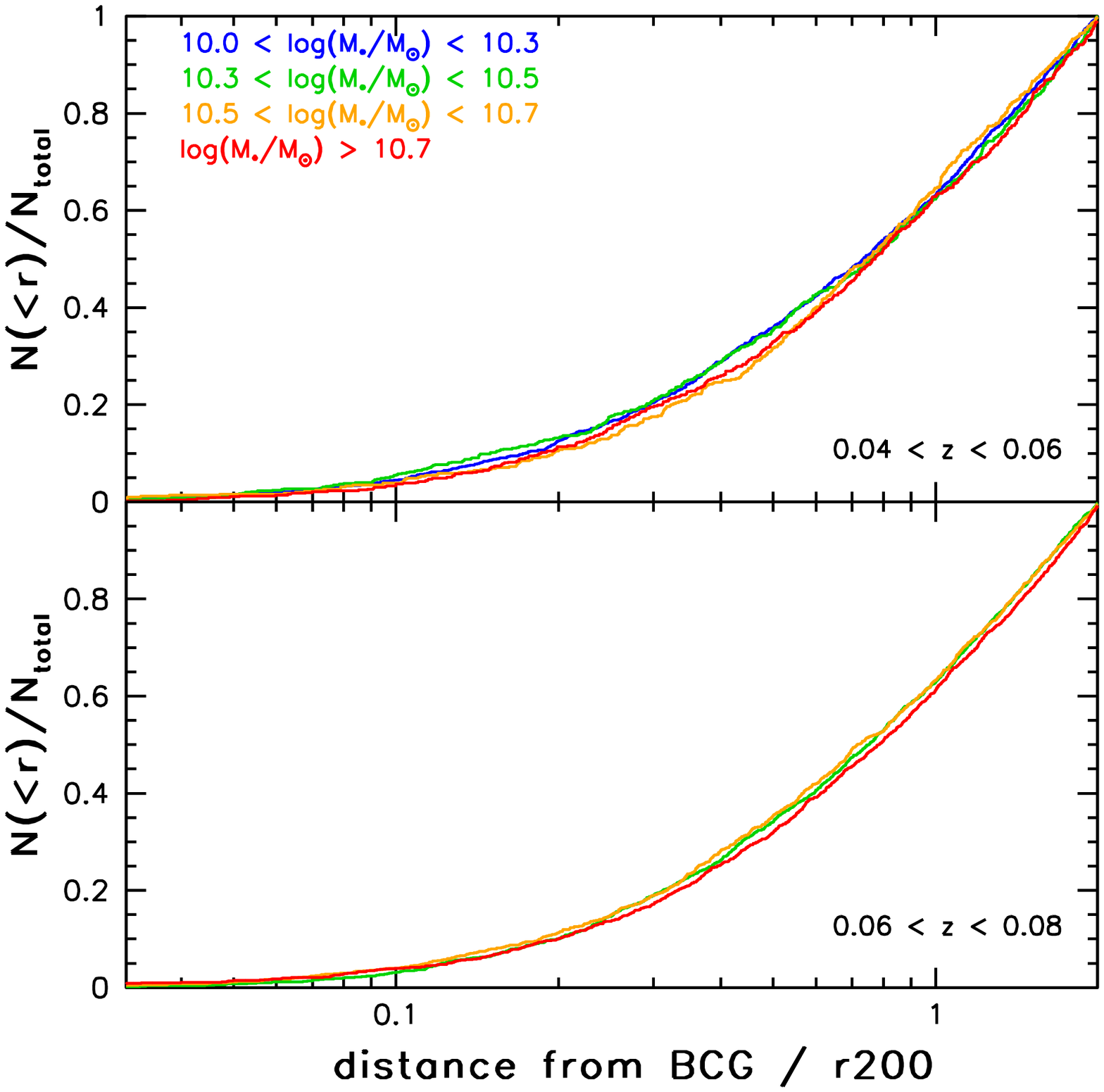}
\caption{~Left: The median mass as function of distance from the BCG
  in the four redshift subsamples. To estimate the significance of
  each data point, we draw 100 bootstrap realizations of the cluster
  sample (rather than the galaxy sample).  The error bars in this and
  the following figures indicate the 68\% confidence intervals of the
  bootstrap. Right: The cumulative radial distributions within
  $2R_{200}$ of four mass bins in two redshift subsamples.  Only the
  results for the two intermediate redshift samples are shown - the
  other two samples yield very similar results. Also note that the
  small offset seen between the two smaller mass bins and the two
  larger mass bins in the $0.04<z<0.06$ redshift samples is
  significant only at the $1\sigma$ level, and is not supported by the
  $0.06<z<0.08$ redshift bin, which has a much larger number of
  objects.}
\label{fig:profiles:mass}
\end{center}
\end{figure*}

\subsection{Mass segregation}
\label{sect:mass}

It has been claimed that the dependence of the galaxy population on
environment is simply a consequence of a varying galaxy stellar mass
function with local density \citep[e.g.][]{bnb08,bpy08}. However,
other studies based on cluster samples find that any luminosity
segregation is solely due to the BCG(s) \citep{bkt02,pdp05}. One would
expect mass segregation in the clusters, if the dynamical friction
timescale is comparable to the cluster crossing time.  The more
massive galaxies then sink to the center faster than less massive
galaxies, thus changing the mass function. In the left panel of
Fig.~\ref{fig:profiles:mass}, we test for this effect by examining the
median mass in the four redshift subsamples. Unlike \citet{bnb08} and
\citet{bpy08}, we find that the median mass appears constant with
clustercentric radius. But the median as function of radius might be
insensitive to changes in the radial distribution, and thus we test
for mass segregation also via the cumulative radial distribution of
four mass subsamples (right panel of Fig.~\ref{fig:profiles:mass}). We
do not combine the redshift subsamples in order to avoid any possible
bias from fiber collisions. If mass segregation were significant, it
should show up as a deficit at small radii for the largest stellar
masses. This is clearly not the case - the radial distributions for
all mass subsamples are very similar. We thus find no indication of
mass segregation in our cluster samples. This likely indicates that
the dynamical friction time in our clusters is larger than the
characteristic cluster crossing timescale for the majority of our
galaxies. From N-body simulations, \citet{bmq08} find that in order
for the dynamical friction time-scale to be of the order of $<5$~Gyr,
the mass ratio between the satellite and the host halo needs to be
large, $\gtrsim 0.1$, and the merger orbit needs to be highly
eccentric. For individual galaxies and small groups infalling into the
cluster, this is not the case, and the typical dynamical friction
timescales are longer, $>10$Gyr. The lack of noticeable mass
segregation is thus fully consistent with such long dynamical friction
timescales.

How can this be reconciled with the results of \citet{bnb08} and
\citet{bpy08}? \citeauthor{bnb08} do not exclude the BCGs from their
study, and their stellar mass trend appears to be in fact largely
driven by the BCGs. \citeauthor{bpy08} do exclude the BCGs, but still
find a significant trend of average stellar mass with radius. However,
they find a strong gradient only for clusters of $M_{180} < 10^{14}
h^{-1} M_{\odot}$. For larger halo masses, the gradient is significant
only at the $1\sigma$ level. Almost half of our clusters fall into
this largest halo mass bin, and because these clusters are typically
richer, they dominate our signal. The remaining discrepancy is thus
not large. It is not clear what causes it - the cluster samples differ
in many aspects. One possibility is that the discrepancy is caused by
the different adaptations of the cluster center: \citeauthor{bpy08}
measure the distance to the luminosity-weighted centroid, rather than
the distance to the BCG. It is thus natural that they see luminous
(and thus massive) galaxies at small radii. But particularly in case
they missed the true BCG due to fiber collisions, or in the presence
of significant substructure, the luminosity-weighted centroid may
differ substantially from the position of our BCG.

\subsection{Star formation--radius relation}
\label{sect:spectro:color-radius}

\begin{figure*}
\begin{center}
\includegraphics[width=0.31\hsize]{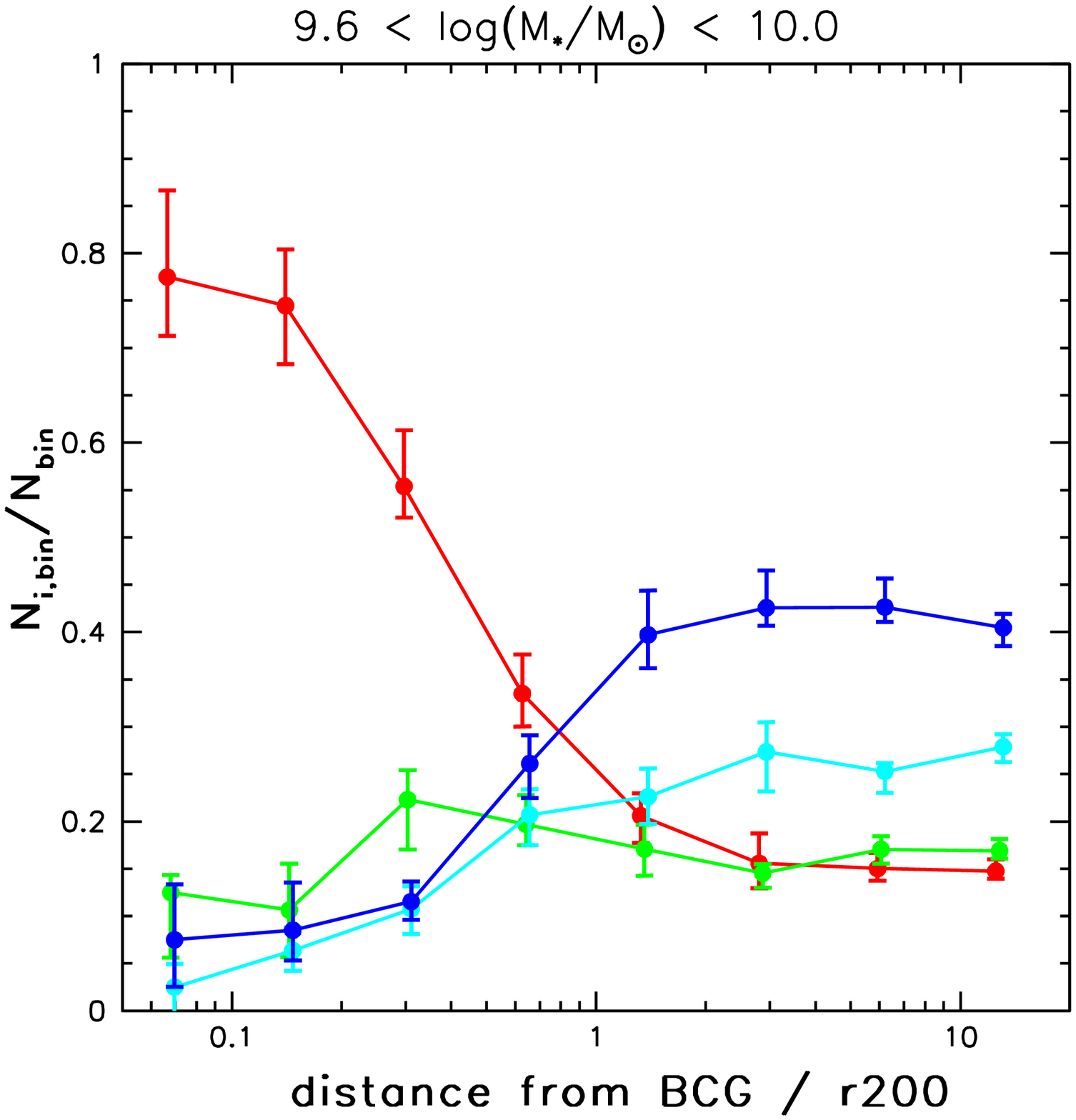}
\includegraphics[width=0.31\hsize]{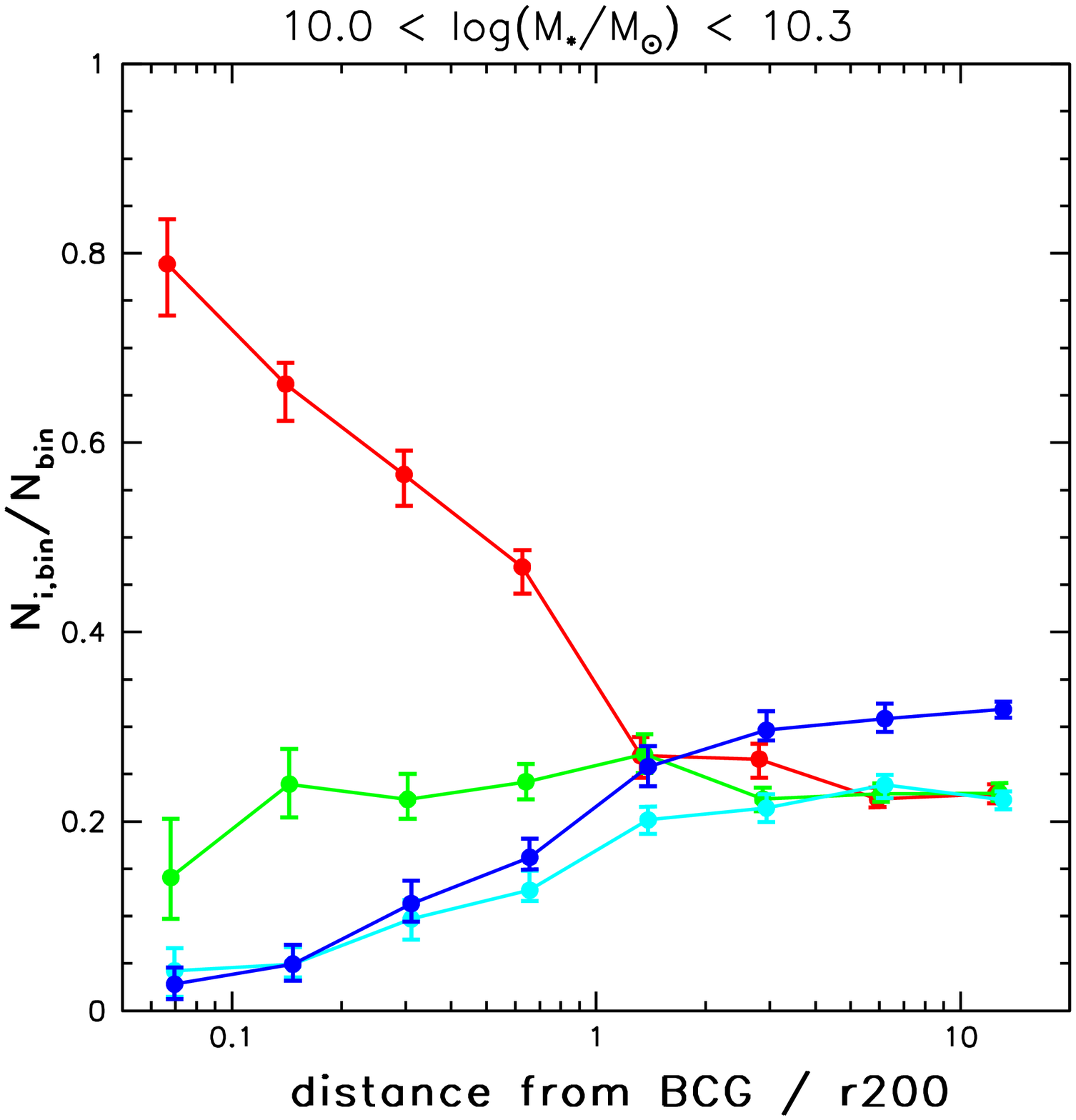}
\includegraphics[width=0.31\hsize]{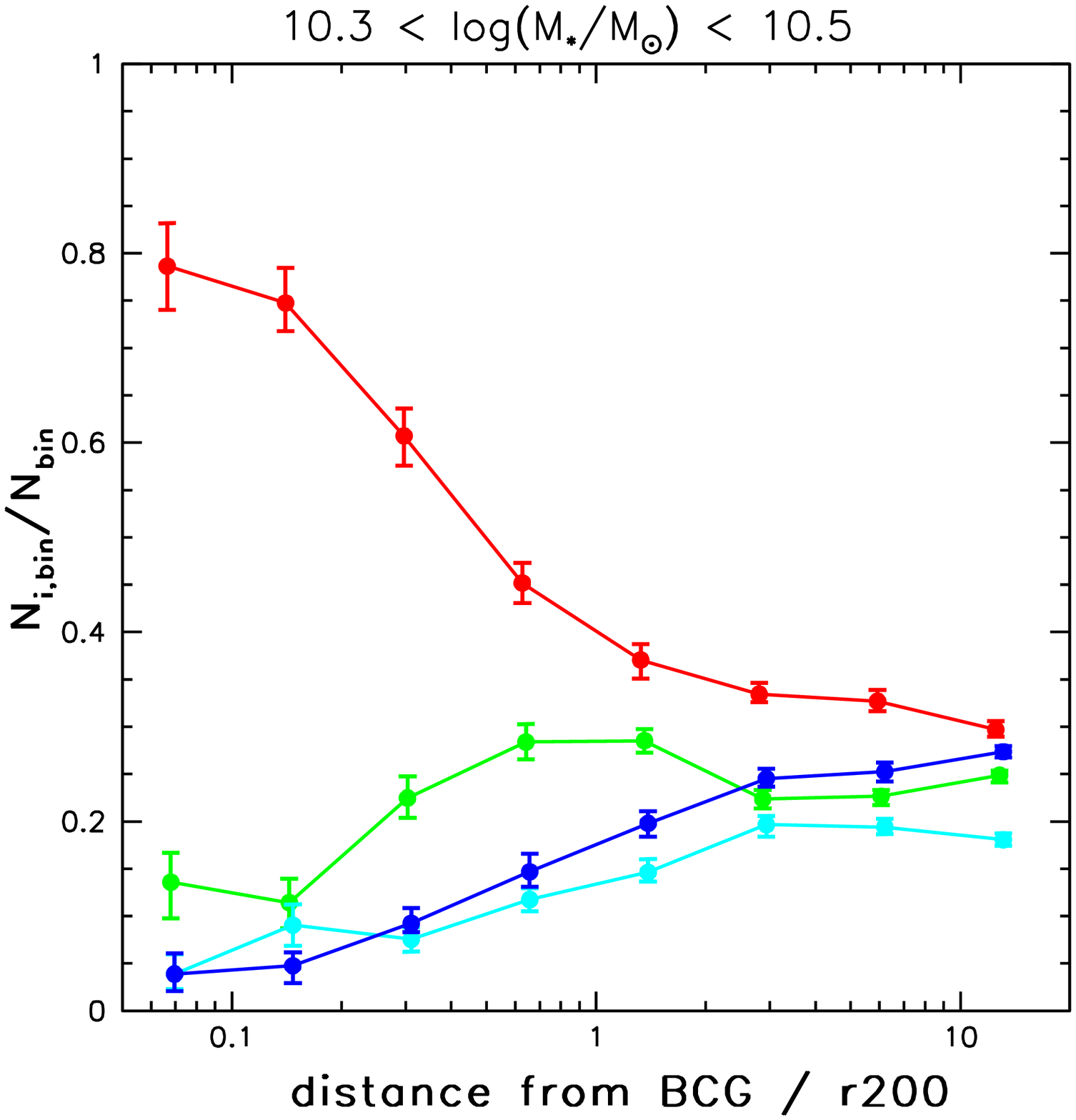}
\begin{minipage}{0.62\hsize}
\includegraphics[width=0.5\hsize]{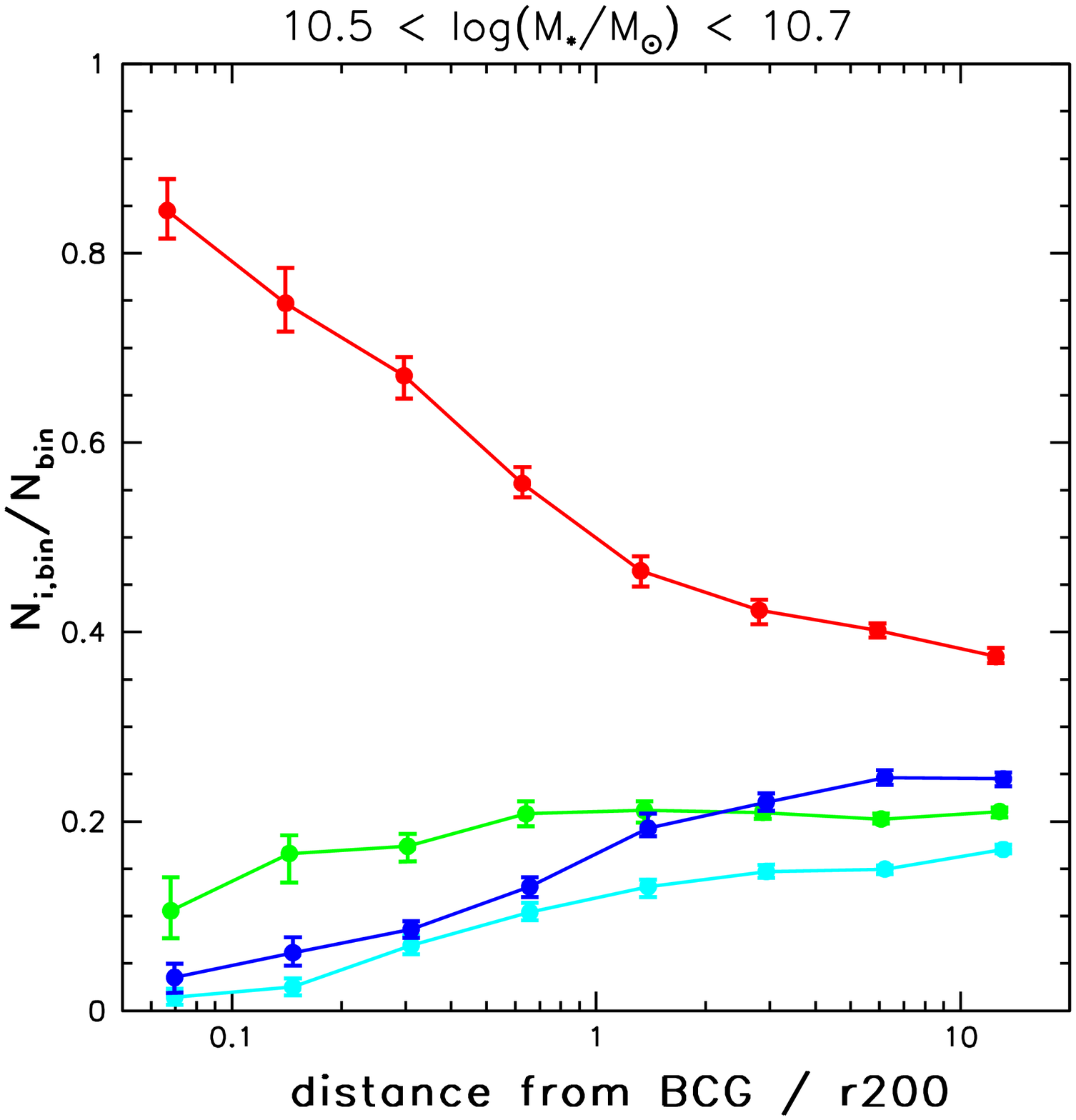}
\includegraphics[width=0.5\hsize]{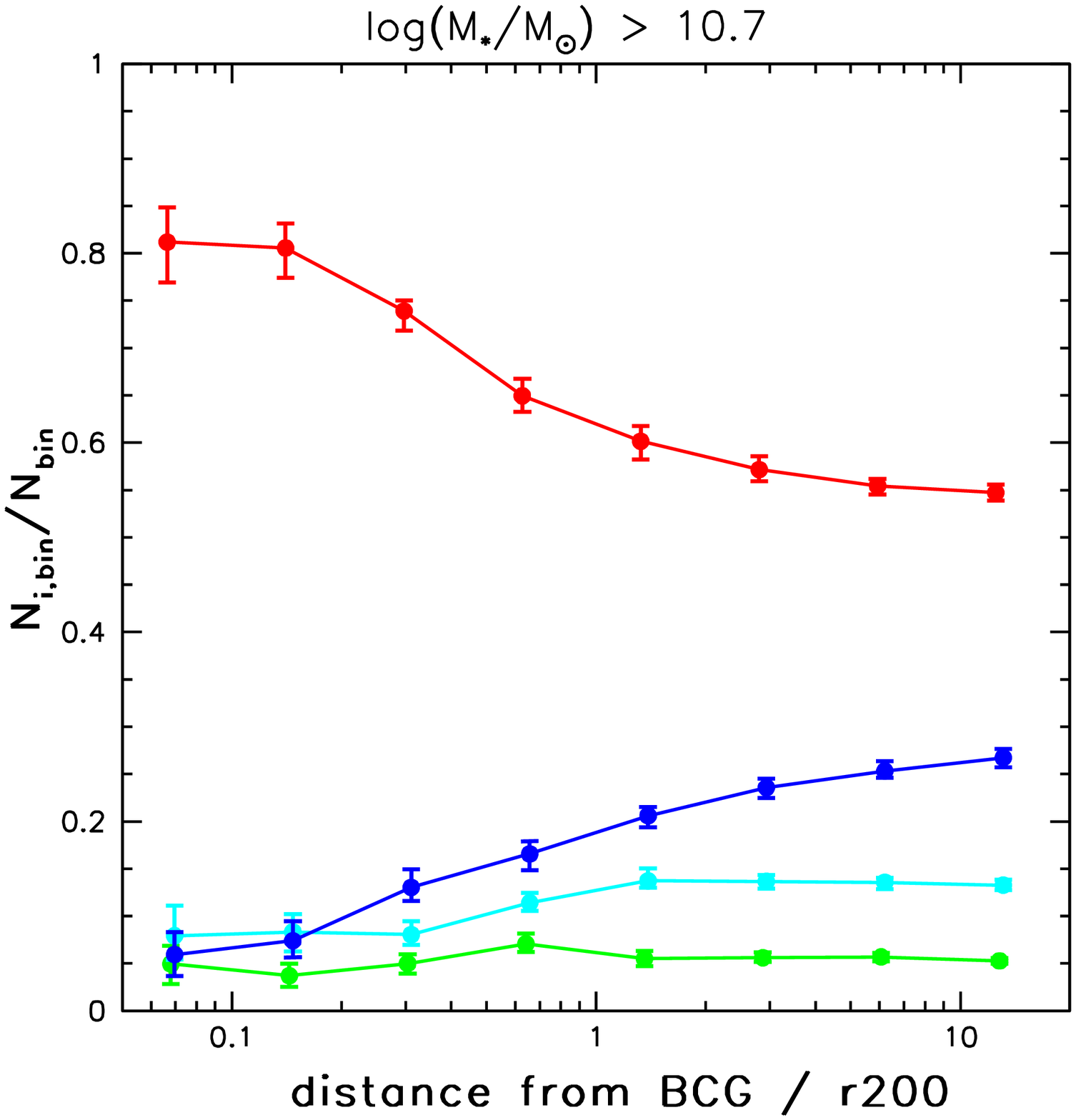}
\end{minipage}
\begin{minipage}{0.31\hsize}
\begin{center}
\begin{minipage}{0.9\hsize}
\caption{~The fractions of red, ``blue'', ``cyan'', and ``green'' galaxies as
  function of distance from the BCG, in five mass ranges (indicated at the
  top of each panel). The color of the
  symbols and their connecting lines indicates the classification.
}
\label{fig:profiles:color_mass_fractions}
\end{minipage}
\end{center}
\end{minipage}
\end{center}
\end{figure*}

In Fig.~\ref{fig:profiles:color_mass_fractions} we investigate how
each galaxy population (red, green, cyan, and blue) contributes to the
number of galaxies within bins of mass and of clustercentric
distance. For each mass range, we use the maximal number of clusters
for which this mass range is complete, i.e. for the lowest mass range,
$9.6<\log(M_{\star}/M_{\odot})<10$, only the lowest-redshift subsample
with $z\le0.04$ is used, for the mass range
$10<\log(M_{\star}/M_{\odot})<10.3$ clusters with $z\le0.06$ are used
(cf.  Fig. \ref{fig:z_mass_samples}). For each mass range, we find a
strong color--radius relation: within about $1R_{200}$, clusters are
dominated by red galaxies. In the center, they make up more than 80\%
of the galaxy population, regardless of mass. The color-radius
relation is particularly pronounced for low mass galaxies: in the
lowest mass range, less than 20\% of the field galaxies are red
galaxies. This fraction is approximately constant till the virial
radius, where it changes abruptly with a steep increase in the red
fraction towards the cluster center. With increasing stellar mass, the
red fraction in the field galaxies also increases, but also here, the
strongest change is between $\sim 0.3 R_{200}$ and $\sim 2
R_{200}$. 

By construction, the majority of field galaxies which are not red
belong to the ``blue'' class.  This is not the case within
clusters. Clearly, the ``blue'' galaxies (i.e. the galaxies with
youngest stellar population ages, and highest star formation rates)
are most strongly affected by the cluster environment. The fraction of
``cyan'' galaxies decreases more gradually. The fraction of ``green''
galaxies, however, is almost constant across all radii.  Only in the
very cluster center ($\lesssim 0.2 R_{200}$) is it significantly lower
than in the field. This is in good agreement with the results of
\citet{wby06}, who found that the fraction of ``intermediate''
galaxies, as classified by $g-r$ color, is independent of environment.

These trends suggest that star formation is affected by a physical
process effective already at $R_{200}$ and beyond. As star formation
shuts off, ``blue'' and ``cyan'' galaxies redden and move onto the red
sequence. This transition apparently does not occur instantaneously,
rather, these galaxies are seen as ``green valley'' objects for a time
span somewhat shorter than a cluster crossing time. Hence the fraction
of green galaxies decreases slower than the number of blue galaxies,
as it is replenished by the latter ones as they redden. Similarly,
``blue'' galaxies will be first seen as ``cyan'', hence the fraction
of the latter declines more slowly.

One striking trend with mass is that the transition in the population
mix appears more abrupt at low masses, and more gradual at high
masses. At low masses, suppression of star formation can traced to
$\sim 2 R_{200}$, but at high masses, up to $\sim 4 R_{200}$.  We have
verified that this is not simply a consequence of cluster
selection effects and the slightly different cluster samples employed
for each mass range. One might conjecture that the more abrupt trend
seen in the low-mass samples is simply a consequence of a possible
higher completeness in selecting smaller galaxy groups at lower
redshifts, and thus a purer field sample at low redshifts. The more
gradual trend seen at higher stellar mass could then be caused by
undetected galaxy groups in filaments around the cluster, which would
already quench star formation prior to cluster infall. To test this
hypothesis, we compared the star formation -- radius relation in a
fixed mass bin, $10.5<\log M_{\odot}<10.7$, in all four redshift
subsamples. The relation is remarkably similar in all four samples,
and so we conclude that the qualitative differences in the mass
subsamples is not caused by a possible redshift dependence of
cluster selection.

It has been suggested that an appreciable fraction of galaxies
observed at $\sim1-3 R_{200}$ previously passed through the cluster
\citep[e.g.][]{bnm00,gkg05,lns09}. This ``backsplash'' population
should be dominated by low-mass, red galaxies. While we see
suppression of star formation beyond $R_{200}$, we see no evidence of
this mass dependence. In all mass bins, the population at the virial
radius and just beyond is very similar to the field. If anything, our
data suggests that more massive galaxies are affected to larger
distances from the cluster.

The apparent influence of the cluster to distances well beyond
$R_{200}$ may be an indication of ``pre-processing'' of galaxies in
the filaments and in small groups below our detection threshold
\citep[e.g.][]{zam98}. With increasing stellar mass, galaxies cluster
more strongly \citep[e.g.][]{lkj06}, and are thus more likely to be
found in such environments. This provides a natural explanation for
the earlier onset of the cluster influence for more massive galaxies.

Finally, the sharp transition from field to cluster serves as evidence
that our measured velocity dispersions and cluster centers, both of
which required manual adjustments to an automatic process, are in
general good representations of their true values.

\begin{figure*}
\begin{center}
\includegraphics[width=0.48\hsize]{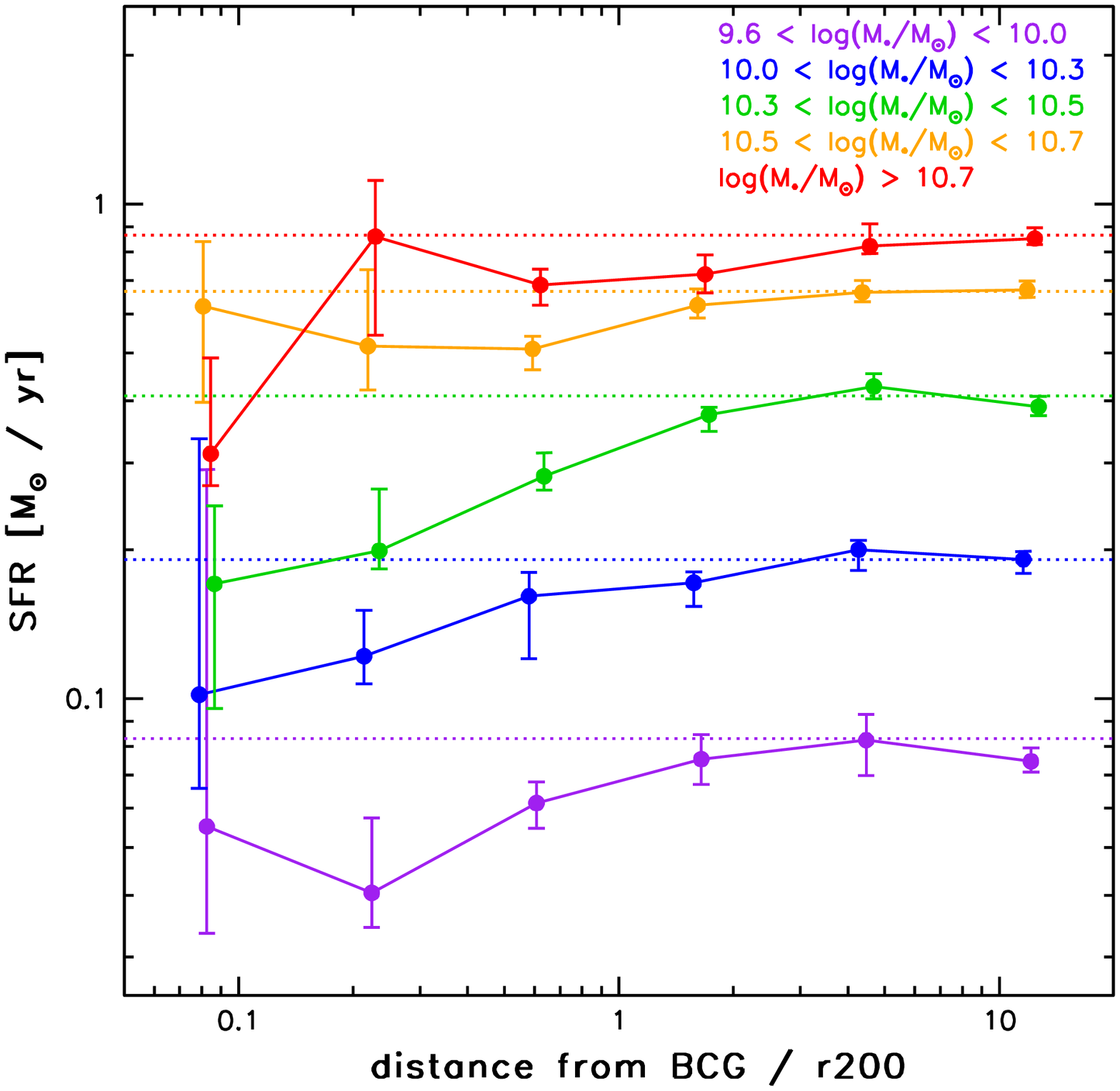}
\includegraphics[width=0.48\hsize]{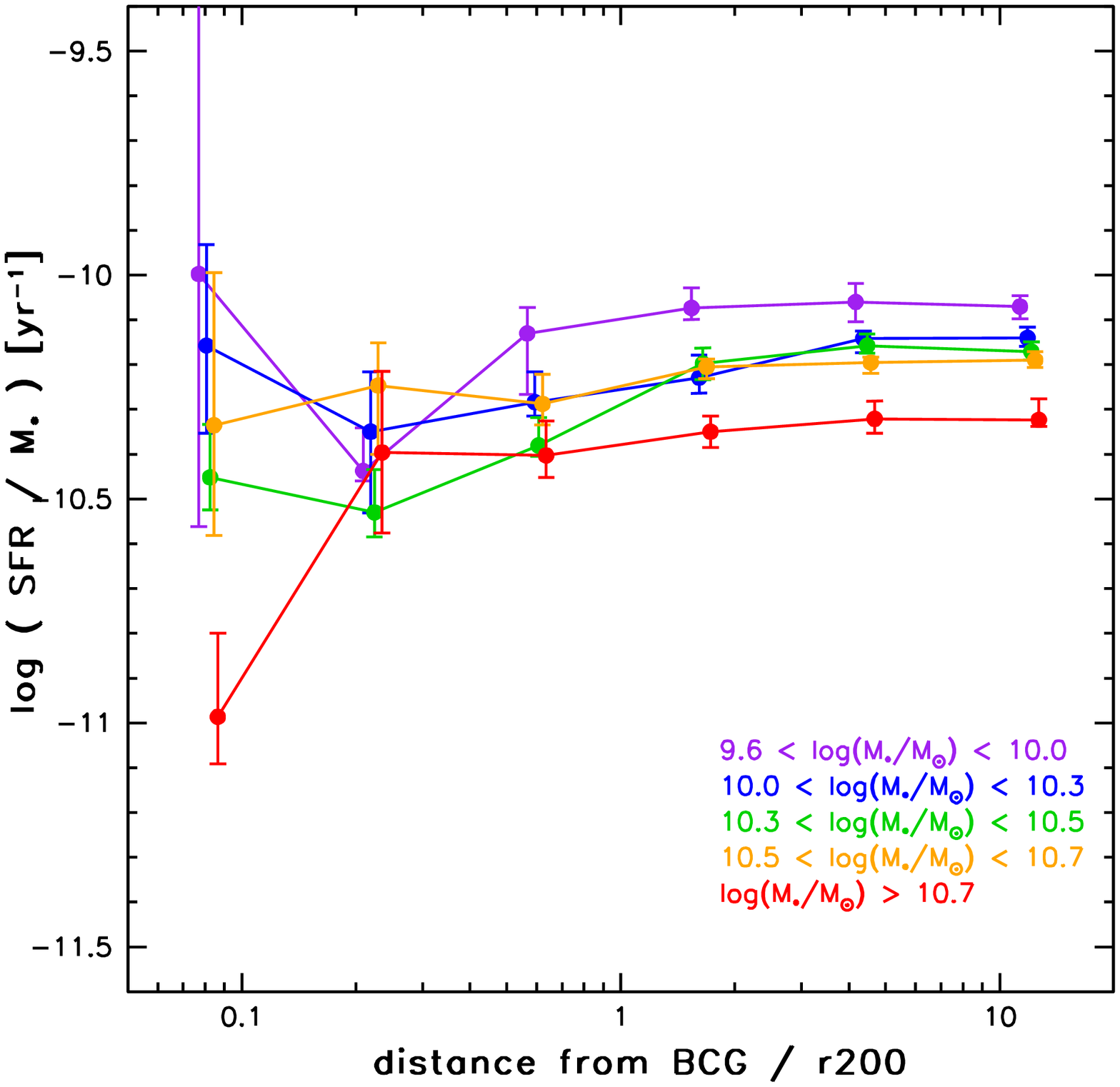}
\caption{~The median star formation rate (left panel) and median
  specific star formation rate (right panel) of ``blue'', ``cyan'',
  and ``green'' galaxies which were classified as star-forming or low S/N
  star-forming galaxies. The dotted lines in the left panel show the
  median value for galaxies beyond the virial radius.  }
\label{fig:profiles:sfr_rate}
\end{center}
\end{figure*}

\subsection{Star formation rate}

The decrease in the fraction of star-forming galaxies means that when
averaging over the whole galaxy population, the star formation rate
clearly declines towards the cluster center. We have shown this via
the ages of the stellar populations of cluster galaxies, which probe
the recent star formation (on timescales $<1$~Gyr). If star formation
indeed declines gradually in cluster galaxies, then also the
instantaneous star formation rates in star-forming galaxies should
decrease towards the cluster center.  Thus, here we investigate the
typical star formation rate of those galaxies with young stellar
population - the ``blue'', ``cyan'', and ``green'' galaxies.

We use the star formation rate measurements of \citet{bcw04}. For
consistency we use only galaxies which have been classified as
star-forming or low S/N star-forming. The star formation rates for
these galaxies are based on emission line model fits to the
spectra. The star formation rates of other types of galaxies (AGN,
composites, and unclassifiable) have been estimated from D(4000) and
are thus representative of the recent star formation history (just as
PC1) rather than the instantaneous rate. Note, however, that our
results do not change if these galaxies are included. Just as the PCA,
the star formation rate measurements are derived only from the central
stellar populations covered by the fiber. We show here the fiber star
formation rates, but our results are qualitatively similar when using
the star formation rates corrected for aperture effects provided by
\citet{bcw04}.

Fig.~\ref{fig:profiles:sfr_rate} shows the median star formation rate
and the specific star formation rate (i.e. the star formation rate
scaled by the galaxy stellar mass) of the star-forming galaxies as
function of cluster distance, again in five mass bins. The star
formation rate clearly decreases towards the cluster, though not very
strongly (less than a factor of 2). This is consistent with the scenario
we proposed earlier: as galaxies fall into the cluster, their star
formation declines. The subsequent aging of the average stellar
population causes the PC1 value of the galaxies to increase - they
transition from ``blue'' via ``cyan'' and ``green'' to ``red''.

\begin{figure}
\begin{center}
\includegraphics[width=0.9\hsize]{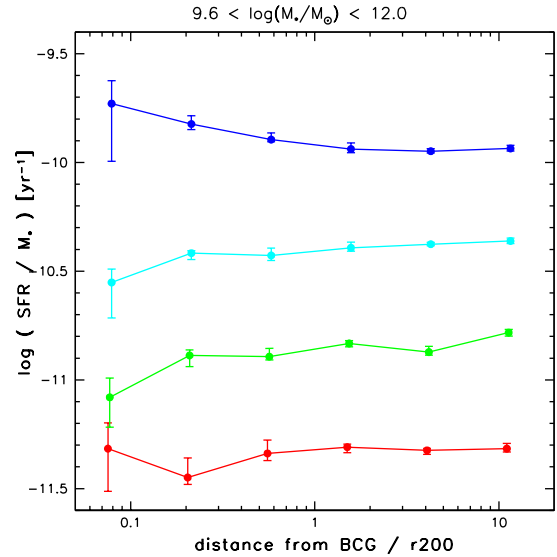}
\caption{~The median specific star formation rate of star-forming
  galaxies, for galaxies of stellar masses
  $\log(M_{\star}/M_{\odot})>9.6$. $\sim 7$\% of ``red'' galaxies are
  classified as (low-SNR) star-forming by \citet{bcw04}.}
\label{fig:profiles:sfr_rate_class}
\end{center}
\end{figure}

In each subpopulation of star-forming galaxies, however, the star
formation rate varies very little with radius
(Fig.~\ref{fig:profiles:sfr_rate_class}) - it is predominantly the
proportions of strongly star-forming (i.e. young, ``blue'') galaxies,
star-forming(``cyan''), and weakly star-forming / transition
(``green'') galaxies which change and cause the decrease in overall
star formation rate.

The gradual decline of the typical star formation rate over most of the
radial range from the cluster outskirts to the cluster center suggests that
the quenching of star formation takes places on timescales similar to the
cluster crossing time, i.e. a few Gyr. Indeed, \citet{bnm00} modeled the
star formation gradients in clusters, assuming that star formation declines
on Gyr timescales, and predicted a decline by a factor of about 3, which is
similar to what we observe.

Our results differ from those of \citet{bem04}, who investigated star
formation rate as a function of environment density in SDSS and 2dFGRS
galaxies. While they also found that the fraction of actively
star-forming galaxies is a function of density, they did not find a
significant difference in the distribution of H$\alpha$ equivalent
widths (as measure of specific star formation) of star-forming
galaxies in different environments. A possible cause for this
difference is that their density estimator, the distance to the fifth
nearest neighbor, is a more noisy density measurement than
clustercentric radius in our composite cluster. Furthermore, a pure
local density estimator lacks a clear association with a time-scale
since infall into a dense environment, which could be the determining
factor.

\subsection{Fraction of Balmer-strong galaxies}
\label{sect:spectro:psb}

We have argued that the radial dependence of the central stellar
populations of cluster galaxies suggests that star formation is not
shut off instantaneously. This hypothesis can be tested via the
occurrence of galaxies with excess Balmer absorption line strength. If
a galaxy experiences a brief starburst, or if a significant level of
star formation is truncated on short timescales ($< 10^6$yr), the
galaxy spectrum displays strong Balmer absorption lines for $\sim
0.5$Gyr. The PCA measures the excess strength of the Balmer lines with
the second Principal Component, PC2 (see
Fig.~\ref{fig:tracks_pc1_pc2_d4_hd}). While the PCA is a powerful tool
to identify true post-starburst galaxies \citep{wwj09}, we are in this
analysis interested simply in the extremes of the star-forming
population.  In each redshift and mass bin, we identify the Balmer
strong galaxies as the 5\% of blue, cyan, or green galaxies with the
largest PC2 values. This cut-off generally lies around PC2$\sim0$.
Since PC2 is more sensitive to noise in the spectra than PC1, this
analysis is carried out only for galaxies with SNR$\ge$8 spectra
($\sim$90\% of the galaxies). Because of the low number of objects, we
split the population only in two mass bins, with the division at the
galaxy bimodality demarcation mass, $10^{10.5}M_{\odot}$.

\begin{figure}
\begin{center}
\includegraphics[width=0.9\hsize]{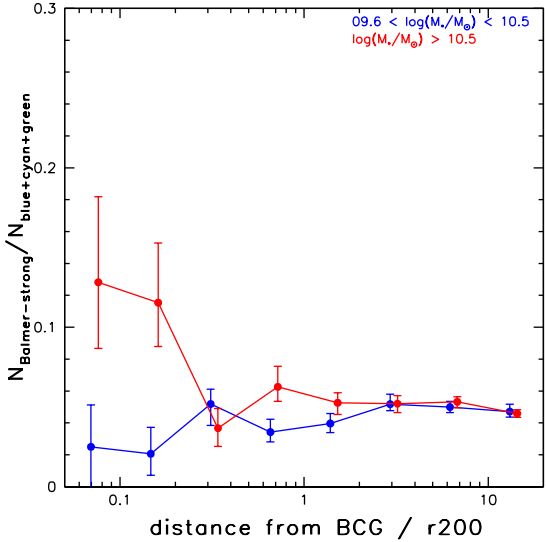}
\caption{~The fraction of Balmer-strong ``blue'', ``cyan'', and ``green''
  galaxies in two mass bins, as function of clustercentric distance.}
\label{fig:profiles:psb_fraction}
\end{center}
\end{figure}

Fig.~\ref{fig:profiles:psb_fraction} shows the fraction of
Balmer-strong blue, cyan, and green galaxies as function of distance
from the cluster center. For low-mass galaxies, the fraction of
Balmer-strong galaxies is consistent with being constant with
radius. For high-mass galaxies, the data suggest a tantalizing
increase in the two innermost bins, at $\lesssim 0.2 R_{200}$. The
signal is robust against variations in the definition of Balmer-strong
galaxies, the signal-to-noise criterion, the mass binning, etc. This
is exactly the kind of signal one would expect from ram-pressure
stripping, and it is tempting to associate it with this. However, the
signal is carried by only 11 Balmer-strong galaxies in the first two
bins. Furthermore, comparing the properties of the host clusters,
these galaxies are found in clusters with similar velocity dispersions
and X-ray luminosities as normal star-forming galaxies, whereas the
efficiency of ram-pressure should scale with cluster mass \citep[but
see][for evidence for Balmer-strong galaxies in low-mass
groups]{paz09}. Finally, the known galaxies in the process of being
ram-pressure stripped tend to be low-mass galaxies, but our data show
no peak for low-mass galaxies. The increase in Balmer-strong high-mass
galaxies at the core may therefore be a statistical fluke.

\citet{hmb06} also found a marginal increase in the number of
post-starburst galaxies compared to the number of star-forming
galaxies within clusters. They find an excess (compared to the field
ratio) over all of the radial range up to $R_{200}$, with the largest
ratios at the cluster core ($<0.2 R_{200}$), as well as right at
$R_{200}$. While we cannot confirm the peak at $R_{200}$, nor an
excess beyond $0.2 R_{200}$, it is interesting that both studies find
a peak excess at the cluster core, though it should be kept in mind
that both are only marginal detections.  

With the possible exception of the cluster core, we find no strong
environmental dependence of the ratio of Balmer-strong galaxies to
star-forming galaxies. Since the fraction of star-forming galaxies
decreases in the cluster, the absolute fraction of Balmer-strong
galaxies also decreases towards the cluster center. This is consistent
with the results of \citet{zzl96} and \citet{hmb06}, who found that
post-starburst galaxies are predominantly field galaxies. In our
terminology, this is simply a reflection of the higher field
abundances of blue, cyan, and green galaxies.

For low-mass galaxies, which are the dominant population of infalling,
star-forming galaxies, the lack of a trend in the occurrence of
post-starburst galaxies with cluster environment shows that the
physical process(es) which cause a post-starburst signature in the
spectrum operate no differently throughout most of the cluster
environment than in low-density environments.  Since a (moderate)
post-starburst signature results not only from a recent star-burst,
but also from any rapid truncation of star formation, this indicates
that the process(es) which quench star formation in these galaxies
operate on
longer timescales.

Given that examples of (low-mass) galaxies in the process of being
ram-pressure stripped have been observed, why do we not detect such a
population statistically? One part of the puzzle is probably that these
objects are quite rare, as evidenced by the few objects known. One may
speculate that it must take unusual circumstances for a small disk
galaxy to travel to the cluster core with its star-forming disk
intact. The other part of the puzzle is likely the selection of
clusters - our sample contains only few clusters of similar mass to
the host clusters of these galaxies, and thus ram-pressure stripping
is likely less efficient in our sample.

\subsection{Fraction of strong Active Galactic Nuclei}

\citet{kht03} showed that powerful ($L\mbox{[O{\sc iii}]}>10^7
L_{\odot}$) narrow-line AGN typically reside in massive galaxies with
young stellar populations, i.e. those galaxies which are the exception
to the typical mass bimodality. Nuclear activity and star formation
must therefore be tightly linked in these galaxies, and AGN could thus
provide us with another channel to study environmental influences.

\begin{figure*}
\begin{center}
\includegraphics[width=0.31\hsize]{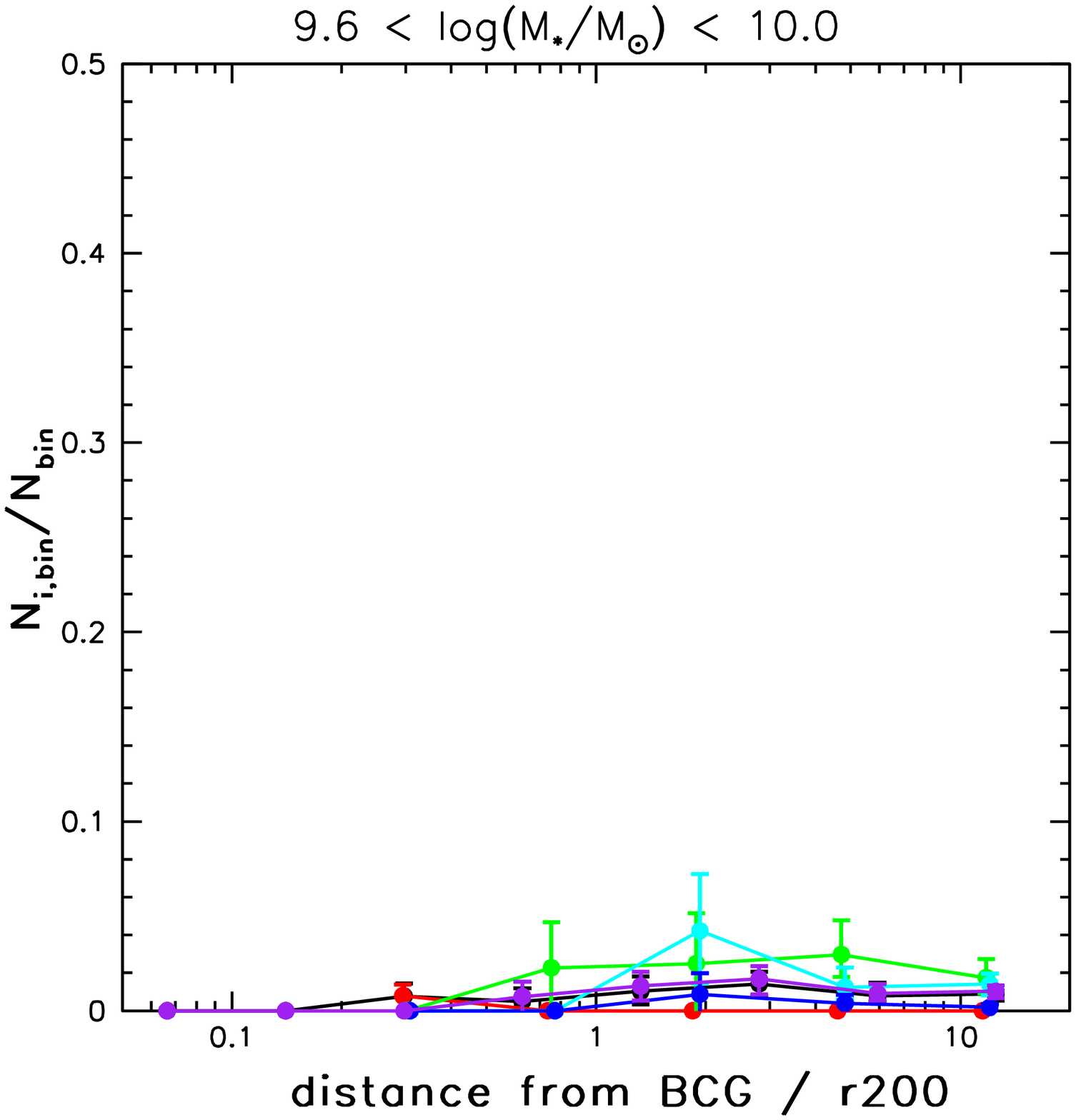}
\includegraphics[width=0.31\hsize]{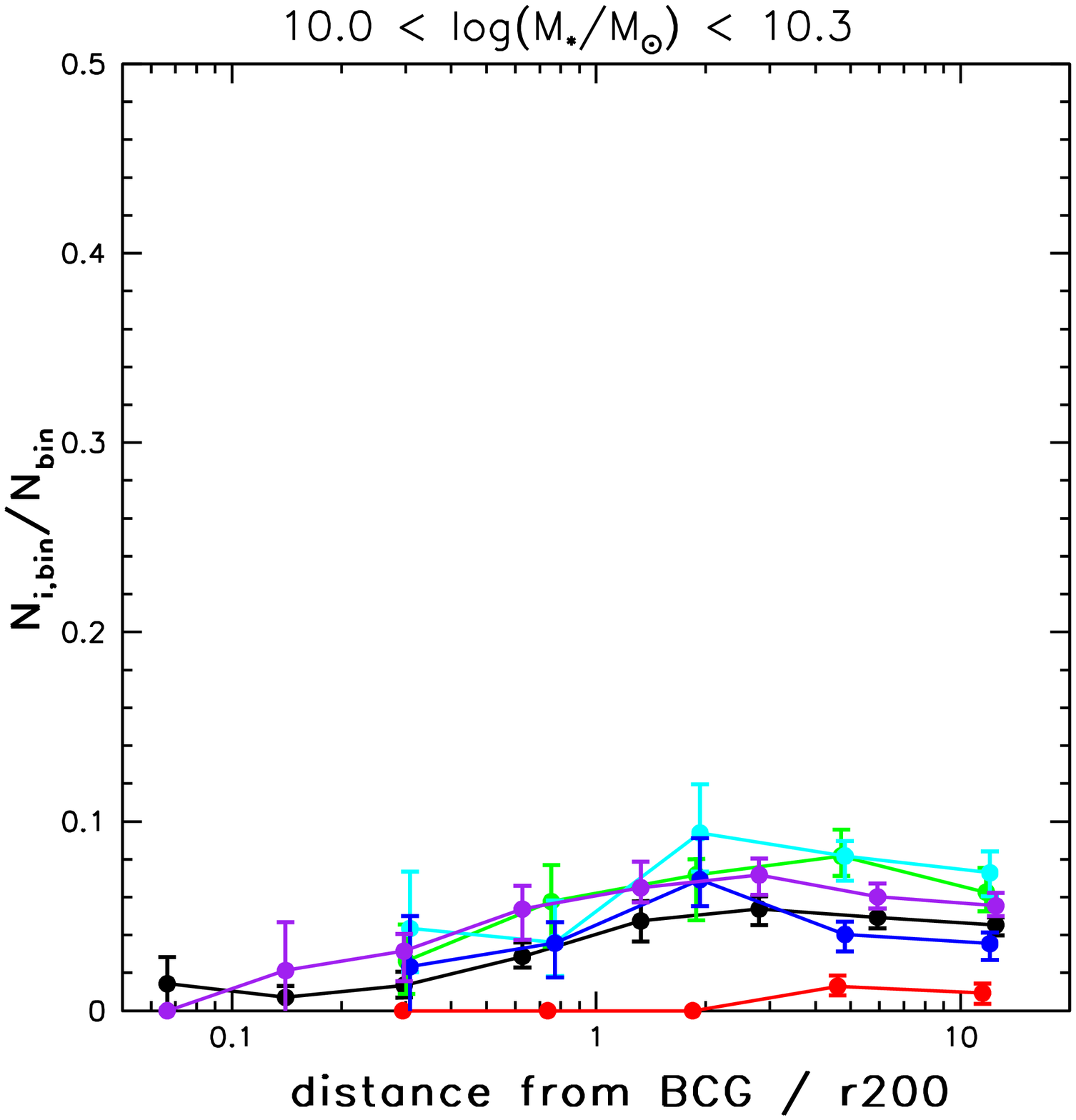}
\includegraphics[width=0.31\hsize]{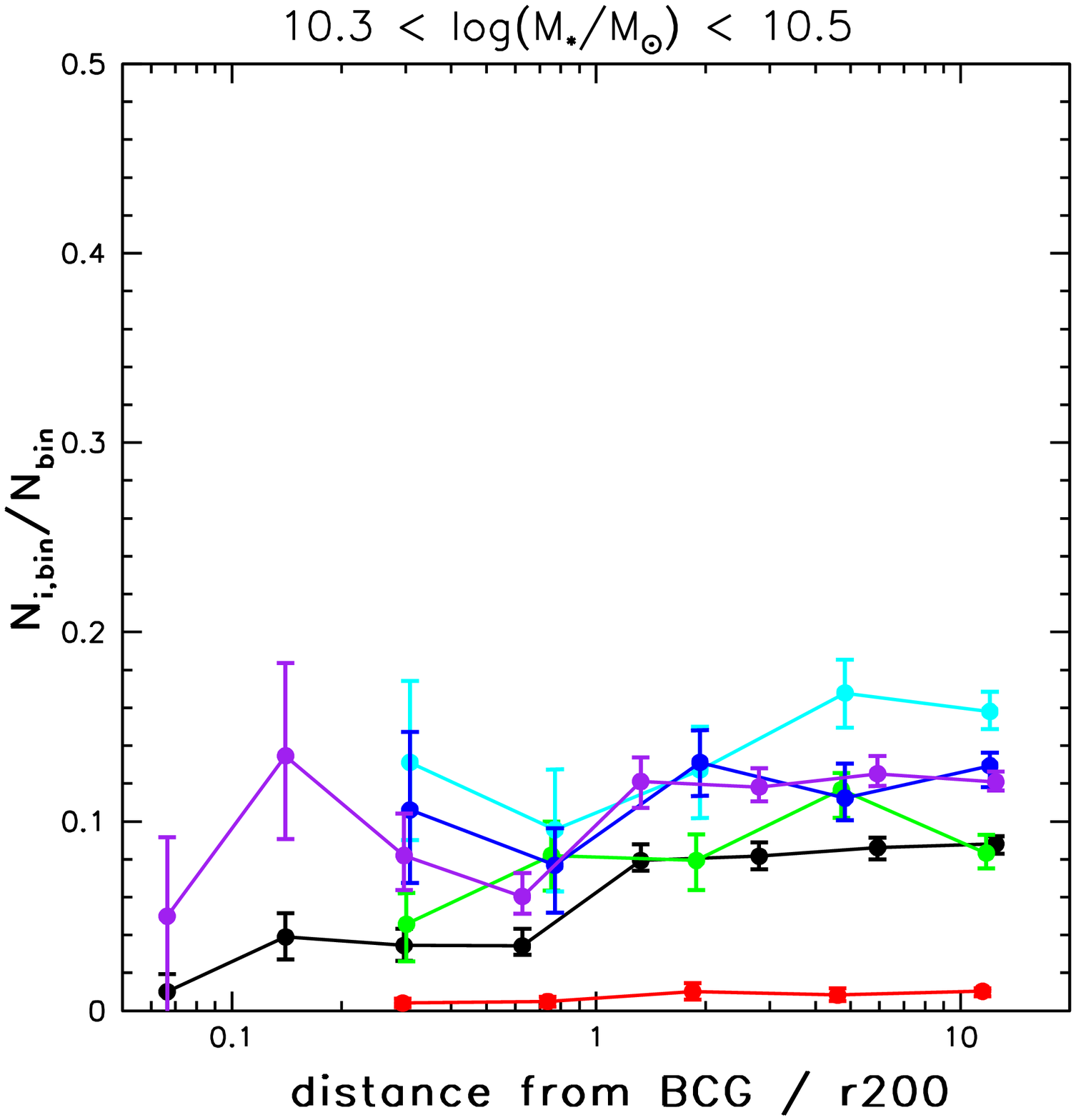}
\begin{minipage}{0.62\hsize}
\includegraphics[width=0.5\hsize]{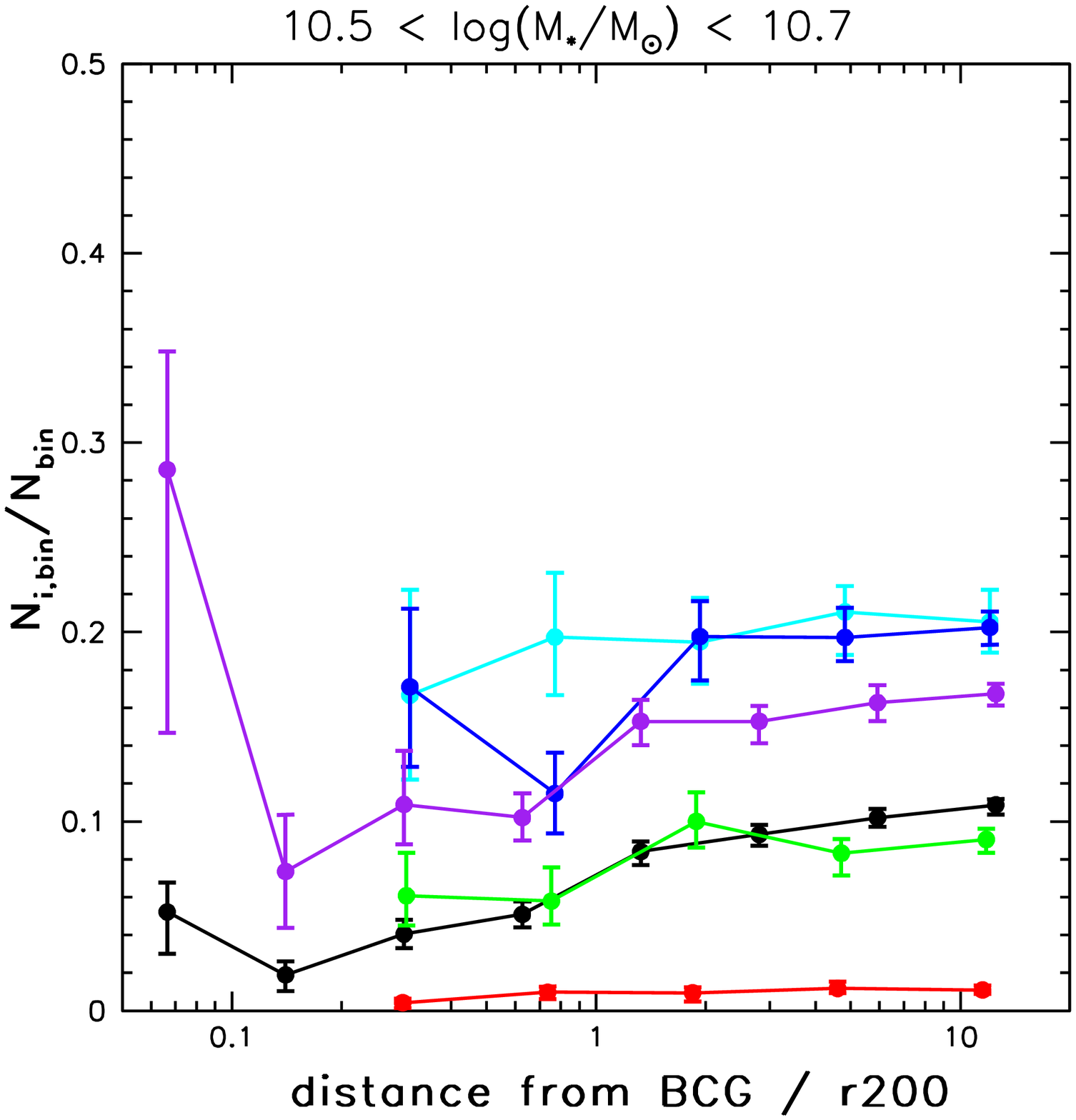}
\includegraphics[width=0.5\hsize]{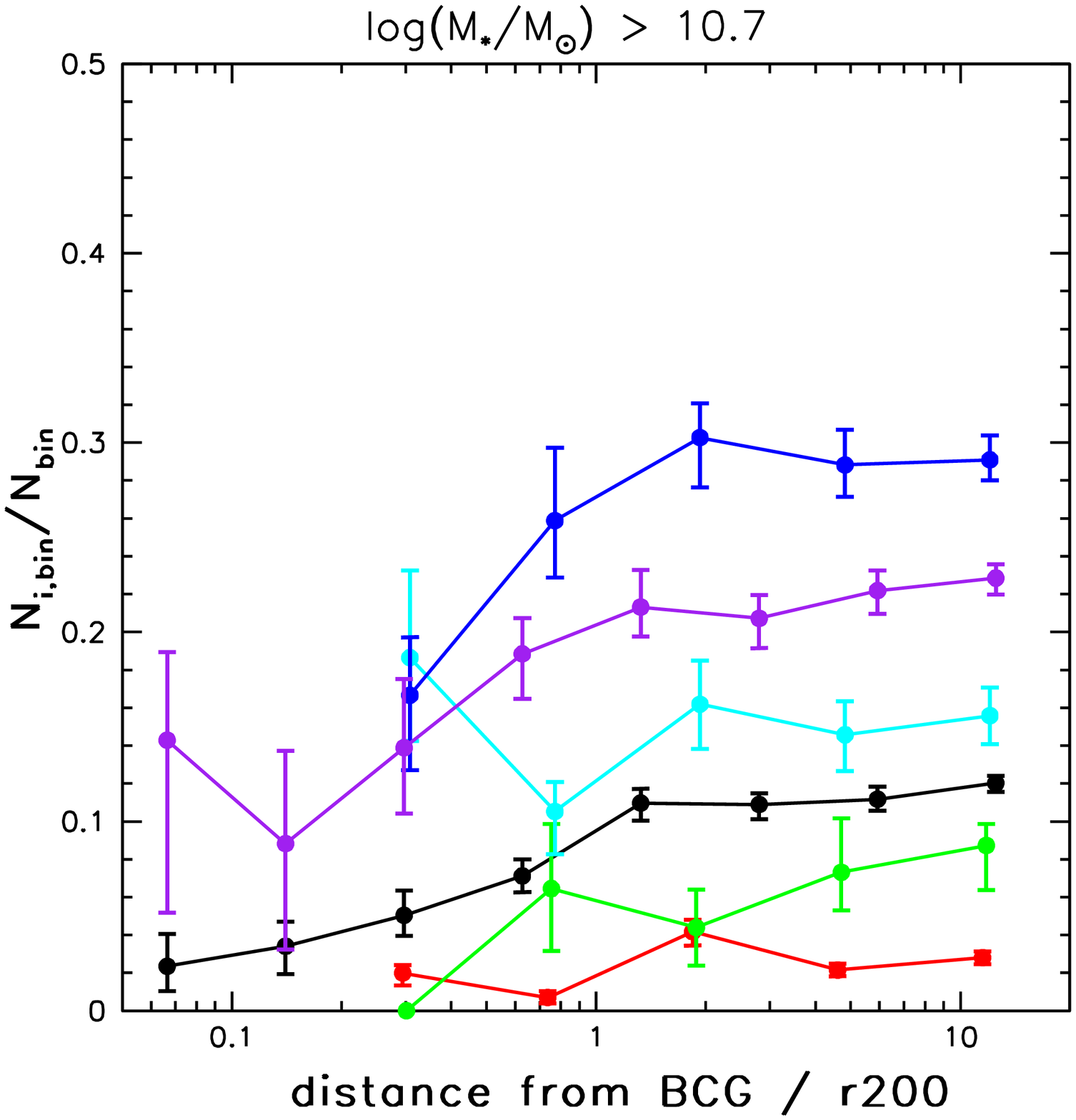}
\end{minipage}
\begin{minipage}{0.31\hsize}
\begin{center}
\begin{minipage}{0.9\hsize}
  \caption{~The fraction of powerful optical AGN ($L\mbox{[O{\sc
        iii}]}>10^7 L_{\odot}$) as function of clustercentric
    distance, in each of the five mass bins. The black dots and lines
    show the overall fraction, the red, green, cyan, and blue lines
    show the fraction of the corresponding galaxy populations. The
    purple symbols show the fraction in ``green'', ``cyan'', and
    ``blue'' galaxies combined.}
\end{minipage}
\end{center}
\label{fig:profiles:strong_agn_fractions}
\end{minipage}
\end{center}
\end{figure*}

A luminosity of $L\mbox{[O{\sc iii}]}>10^7 L_{\odot}$ is approximately
the demarcation boundary between powerful Seyfert galaxies, and
weaker LINERs. Furthermore, above this luminosity, AGN can be reliably
identified in SDSS spectra regardless of redshift and
star formation rate of the host \citep{kht03}.

Fig.~\ref{fig:profiles:strong_agn_fractions} shows the fractions of
galaxies which host a powerful optical AGN with $L\mbox{[O{\sc
    iii}]}>10^7 L_{\odot}$.  We see that the overall fraction of
strong AGNs decreases towards the cluster center. Within each galaxy
class, however, the fraction of strong AGN is consistent with being
constant with radius, at least for radii $\gtrsim 0.2 R_{200}$. The
decline of the total AGN fraction is then almost exclusively an effect
of the increasing number of red galaxies, which host very few strong
AGN.

To increase the number statistics in the core of the cluster, we
investigate the fraction of strong AGN in blue, cyan, and green
galaxies combined (the purple symbols in
Fig.~\ref{fig:profiles:strong_agn_fractions}). We do not find
significant evidence for an enhanced fraction of strong AGN even in
the cluster core.

\citet{lkh08} and \citet{rhr09} showed that kinematical disturbances
such as close companions, mergers, and disk lopsidedness trigger star
formation at the centers of galaxies. While they also find an
increased occurrence of optical AGN, this is entirely accounted for by
the existence of the young stellar populations. Also \citet{kah09}
argue that for the powerful AGN hosted by star-forming galaxies, black
hole accretion rate is largely independent of processes in the
interstellar medium, as long as the central stellar population is
young. These studies suggest that the star formation rate at the
galaxy center regulates the probability that the central black hole
shines as a powerful optical AGN.  Our results are consistent with
this scenario: our galaxy classification scheme is based on the age of
the central stellar population, and thus the AGN fraction within each
galaxy class is independent of cluster radius. Due to the suppression
of star formation, the overall AGN population is suppressed in the
cluster environment, as well.
Strictly speaking, this scenario precludes the use of powerful AGN as
tracers of environmental processes.

\subsection{Fraction of AGN in red galaxies}
\label{sect:agn}

\begin{figure}
\begin{center}
\includegraphics[width=0.9\hsize]{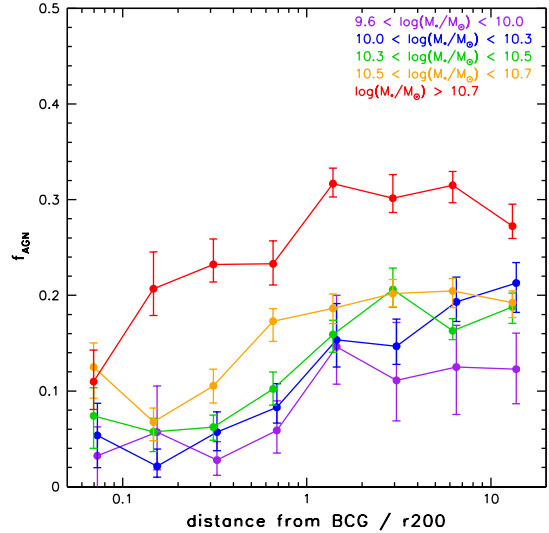}
\caption{~The fraction of red galaxies which host an optical AGN.
}
\label{fig:profiles:red_agn_fraction}
\end{center}
\end{figure}

While the more powerful Seyferts are found almost exclusively in
star-forming galaxies, the weaker LINERs are found predominantly in
galaxies with old stellar populations \citep{kht03}. The star
formation rate in our red galaxies is quite low
(Fig.~\ref{fig:profiles:sfr_rate_class}), so that AGN can be
identified to much lower luminosities without contamination from star
formation emission lines. Fig.~\ref{fig:profiles:red_agn_fraction}
shows that the fraction of optical AGN in red galaxies declines
towards the cluster center, by approximately a factor of two between
the field and the center. The decline is gradual, and consistent with
setting in around $R_{200}$.

The relation between these weak AGN and their host galaxy properties
is very different from the more powerful optical AGN. \citet{kah09}
showed that for these systems, the rate of accretion onto the black
hole is proportional to the bulge mass, with a constant of
proportionality that depends on the age of the bulge stellar
population. A likely explanation for this result is that the AGN is
fuelled by mass loss from evolved stars.  D$_n$(4000) and PC1 are
largely insensitive to stellar populations of $\gtrsim2$~Gyr, and thus
our class of red galaxies is far from a homogeneous population. It is
therefore plausible \citep[and indeed expected, see][]{deb06} that the mean
age of the red galaxies increases towards the cluster center, and
provides less fuel for the central black hole.

There is furthermore evidence that the (weak) AGN activity in galaxies
with old central stellar populations may be sensitive to processes in
the outer galaxy regions. \citet{khb06} find that galaxies with an old
central stellar population which host an optical AGN span a wide range
in UV-optical color. In contrast, galaxies with old central
populations, but no AGN emission, almost all have red UV-optical
colors. They interpret the UV emission as residual small levels of
star formation in an outer disk, traceable only via ultraviolet
radiation. These star-forming outer disks presumably trace a
(low-mass) reservoir of cold gas, which is available to fuel star
formation and AGN activity.\footnote{Note, however, that this is not a
  linear relation. Galaxies with UV-blue outer disks span a wide range
  in central star formation rate, age of the central stellar
  population, and black hole accretion rate.}  Given that we find that
fewer red cluster galaxies host an AGN, we expect that the fraction of
red galaxies with a UV-bright outer disk also decreases in the
cluster. Just as star formation in young galaxies is suppressed upon
infall into the cluster environment, also this ``hidden'', residual
star formation in otherwise old galaxies is quenched, presumably
because potential fuel is removed from the outer regions.

\section{Summary and discussion}
\label{sect:conclusion}

Our results can be summarized as follows:
\begin{itemize}
\item The median mass of cluster galaxies other than the BCG does not
  vary with cluster radius. Environmental differences in the galaxy
  population are therefore not solely a result of a varying stellar
  mass function.
\item There is a pronounced relation between distance from the cluster
  center and the composition of the galaxy population. In the cluster
  center, star formation has ceased in $\sim$80\% of the cluster
  galaxies, regardless of stellar mass. The fraction of the youngest
  (``blue'') stellar populations declines most rapidly towards the
  cluster center, less rapidly for slightly older (``cyan'') stellar
  populations, and is almost constant for the transition ``green
  valley'' galaxies.
\item The star formation rates of galaxies which are not on the red
  sequence declines by about a factor of two from the field to the
  cluster center.
\item The fraction of ``blue'', ``cyan'', and ``green'' galaxies which
  have excess Balmer absorption line strength, indicative of a
  post-starburst or truncation of star formation, is independent of
  cluster radius at least for $>0.2 R_{200}$. There is a marginally
  significant increase of Balmer-strong galaxies within $0.2 R_{200}$
  for galaxy masses $>10^{10.5} M_{\odot}$.
\item The fraction of star-forming and transition galaxies which host a
  powerful optical AGN ($L\mbox{[O{\sc iii}]}>10^7 L_{\odot}$) is
  independent of cluster radius.
\item The fraction of quiescent (red) galaxies which host a (weak)
  optical AGN decreases towards the cluster center.
\end{itemize}

Our results indicate that the star formation rate of galaxies
infalling into clusters decreases gradually over a timescale similar
to the cluster crossing time, i.e. a few Gyr. E.g. a simple
approximately exponential decline of the star formation rate suffices
to explain the declining fraction of very young (i.e. strongly star-forming)
in the cluster. These ``blue'' galaxies would then be seen as
star-forming, ``cyan'' galaxies and later as transition,``green''
galaxies, which naturally explains the slower decline of these two
types.

Remarkably, the properties of star-forming and transition galaxies, at
fixed stellar mass and fixed light-weighted age, are largely
independent of clustercentric radius. The star formation rate in each
of these classes is nearly constant, as well as the fraction of
galaxies which host a post-starburst galaxy or a strong optical AGN.
This argues against more violent processes than the mere fading of
star formation being a dominant effect: truncation of star formation
due to e.g. ram-pressure stripping would lead to an increase of
Balmer-strong galaxies, whereas mergers and kinematic instabilities
(caused e.g. by harassment) would lead to an increase of starburst
galaxies and Balmer-strong galaxies, and possibly AGN activity. There
is no evidence of any of these over the radial range we can probe,
apart from a marginally significant excess of Balmer-strong galaxies
the cluster core.  

We therefore come to the conclusion that a simple gradual decline of
star formation suffices to explain our results, and that other, more
abrupt, processes cannot play a major role at cluster radii beyond a
tenth of the virial radius, at least not in the moderately massive
clusters we probe here.

Our data alone can only approximate the timescale on which star
formation is quenched in clusters - it must be of the order of the
cluster crossing time, a few Gyr. In Weinmann et al., in prep. we
will compare a similar dataset to semi-analytical models in
order to investigate the timescale more precisely.

\section*{Acknowledgments}

Funding for the Sloan Digital Sky Survey (SDSS) has been provided by
the Alfred P. Sloan Foundation, the Participating Institutions, the
National Aeronautics and Space Administration, the National Science
Foundation, the U.S. Department of Energy, the Japanese
Monbukagakusho, and the Max Planck Society. The SDSS Web site is
http://www.sdss.org/.

The SDSS is managed by the Astrophysical Research Consortium (ARC) for
the Participating Institutions. The Participating Institutions are The
University of Chicago, Fermilab, the Institute for Advanced Study, the
Japan Participation Group, The Johns Hopkins University, the Korean
Scientist Group, Los Alamos National Laboratory, the
Max-Planck-Institute for Astronomy (MPIA), the Max-Planck-Institute
for Astrophysics (MPA), New Mexico State University, University of
Pittsburgh, University of Portsmouth, Princeton University, the United
States Naval Observatory, and the University of Washington.

\bibliography{references.bib}

\label{lastpage}

\end{document}